\documentstyle[eqsecnum,aps]{revtex}

\input{epsf.tex}
\begin{document}
\draft
\twocolumn[\hsize\textwidth\columnwidth\hsize\csname
@twocolumnfalse\endcsname
\preprint{HEP/123-qed}
\renewcommand{\thefootnote}{\alph{footnote}} 
\title{Chaos and the continuum limit in the gravitational $N$-body problem
II. Nonintegrable potentials}
\author{Ioannis V. Sideris\footnote{Electronic address: sideris@astro.ufl.edu}}
\address{Department of Astronomy, University of Florida, Gainesville, 
Florida 32611}
\author{Henry E. Kandrup\footnote{Electronic address: kandrup@astro.ufl.edu}}
\address{ Department of Astronomy, Department of Physics, and
Institute for Fundamental Theory
\\
University of Florida, Gainesville, Florida 32611\\}

\date{\today}
\maketitle
\begin{abstract}
This paper continues a numerical investigation of the statistical properties
of orbits evolved in `frozen,' time-independent $N$-body realisations of 
smooth time-independent density distributions ${\rho}$ 
corresponding to both integrable and nonintegrable potentials, allowing for 
$10^{2.5}{\;}{\le}{\;}N{\;}{\le}{\;}10^{5.5}$. 
The principal focus is on distinguishing between, and quantifying, 
the effects of graininess on initial conditions corresponding, in the 
continuum limit, to regular and chaotic orbits. Ordinary Lyapunov exponents
${\chi}$ do not provide a useful diagnostic for distinguishing between
regular and chaotic behaviour. Frozen-$N$ orbits corresponding in the
continuum limit to both regular and chaotic characteristics have large 
positive ${\chi}$ even though, 
for large $N$, the `regular' frozen-$N$ orbits closely resemble regular 
characteristics in the smooth potential. Alternatively, viewed macroscopically 
both `regular' and `chaotic' frozen-$N$ orbits diverge as a power law in time 
from smooth orbits with the same initial condition. There {\em does}, however,
exist an important difference between `regular' and `chaotic' frozen-$N$
orbits: For regular orbits, the time scale associated with this divergence
$t_{G}{\;}{\sim}{\;}N^{1/2}t_{D}$, with $t_{D}$ a characteristic dynamical, or
crossing time; for chaotic orbits $t_{G}{\;}{\sim}{\;}(\ln N)t_{D}$. 
Convergence towards the continuum limit is much slower for chaotic
orbits. At least for $N>10^{3}$ or so, clear distinctions exist between phase 
mixing of initially localised orbit ensembles which, in the continuum limit,
exhibit regular versus chaotic behaviour. Regular ensembles evolved in a 
frozen-$N$ density distribution diverge as a power law in time, albeit more 
rapidly than ensembles evolved in the smooth distribution. Alternatively, 
chaotic ensembles diverge in a fashion that is roughly exponential, albeit at
a larger rate than that associated with the exponential divergence of the
corresponding ensemble evolved in the smooth ${\rho}$. For both regular and
chaotic ensembles, finite-$N$ effects are well mimicked, both qualitatively
and quantitatively, by energy-conserving white noise with amplitude 
${\eta}{\;}{\propto}{\;}1/N$. 
This suggests strongly that earlier investigations of the effects of low
amplitude noise on phase space transport in smooth potentials are directly
relevant to real physical systems.

\end{abstract}
]
\narrowtext
\section{MOTIVATIONS AND EXPECTATIONS}
 \label{sec:level1}
This is the second in a series of papers, the aim of which is to understand
the role of discreteness effects, {\em i.e.,} graininess, in the gravitational
$N$-body problem. Particular emphasis is given to the meaning of chaos and 
various manifestations of chaotic behaviour. As discussed in the first 
paper\cite{KS} (hereafter Paper I), this problem can be divided into two 
separate components, namely first understanding how graininess alters the 
motions of representative orbits in a fixed gravitational potential and only
then considering how these changes are manifested in the context of a fully 
self-consistent $N$-body evolution. As in Paper I, the focus here is on the 
former issue. 

One is thus led naturally to effect a statistical comparison between
(1) orbits evolved in a frozen $N$-body density distribution generated by 
randomly sampling some specified smooth density distribution ${\rho}$ and (2) 
orbits evolved in the smooth potential ${\Phi}$ related to ${\rho}$ by 
Poisson's equation, ${\nabla}^{2}{\Phi}=4{\pi}G{\rho}$. 

In this setting, the present paper has two specific objectives, namely (1)
to implement precise, quantitative distinctions between `regular' and `chaotic'
behaviour in such frozen-$N$ systems; and (2) to determine the extent to which
discreteness effects can be mimicked successfully by a suitably defined `noise'
acting on orbits evolving in an otherwise smooth potential.

Implementing useful distinctions between regular and chaotic behaviour is
not completely trivial. For example, ordinary Lyapunov exponents computed
for individual orbits do not provide a useful diagnostic. Even for 
density distributions corresponding to integrable potentials, $N$-body orbits 
have large positive Lyapunov exponents. Moreover, even though there is a 
precise sense in which, as $N$ increases, frozen-$N$ orbits come to more 
closely resemble orbits in the smooth potential, the values of these exponents 
do {\em not} decrease systematically with increasing $N$ \cite{VM}\cite{KS}.

Viewed macroscopically, a frozen-$N$ orbit and a smooth orbit evolved from the
same initial condition in density distributions corresponding to an
integrable potential will typically diverge {\em linearly} in time on a
time scale $t_{G}{\;}{\propto}{\;}N^{1/2}$. As discussed 
in Paper I, this superficially surprising result would appear to reflect 
the fact that the chaos is associated with a large number of encounters with
neighbouring particles, each of very short duration, which tend to cancel
systematically so as to have a comparatively minor macroscopic effect.

For frozen-$N$ systems with very small $N$, one would anticipate that
close encounters are relatively stronger and `more {\em macro}scopic' in nature
than is the case for larger $N$, so that clean macroscopic distinctions 
between regular and chaotic behaviour might be difficult to implement. 
However, for larger $N$ such close encounters become progressively `more 
{\em micro}scopic', so that one might expect comparatively clear cut 
distinctions to exist. 

Following the pioneering work of 
Chandrasekhar in the 1940's\cite{Ch}\cite{Ch2}, one 
might expect that discreteness effects will act in much the same way as
friction and white noise, so that they can be modelled in the context of
a Fokker-Planck description. To the extent that this intuition is correct,
earlier work probing the effects of friction and noise on smooth potential
orbits translate into precise predictions as to the expected effects of
graininess.

The amplitude ${\eta}$ associated with friction and noise defines a 
characteristic relaxation time $t_{R}=1/{\eta}$ on which, {\em e.g.,} the
perturbation will induce significant changes in conserved quantities like 
energy. However, modelling discreteness effects as a sequence of close binary
encounters which result in friction and noise yields the concrete 
prediction\cite{Ch2} that $t_{R}{\;}{\propto}{\;}(N/ \ln N)t_{D}$, where 
$t_{D}$ is a characteristic dynamical, or crossing, time. One might, 
therefore, anticipate that discreteness effects associated with a system 
comprised of $N$ bodies can be reproduced by friction and noise with 
characteristic amplitude ${\eta}{\;}{\propto}{\;}\ln N/N$. 

It is well known that, when subjected to friction and white noise, orbits in 
an integrable potential and regular orbits in a generic potential typically 
diverge as a power law from unperturbed orbits with the same initial condition.
One might, therefore, expect that discreteness effects also induce a power
law divergence between frozen-$N$ orbits and smooth characteristics with
the same initial condition, and that the divergence time scale $t_{G}$ 
associated with a system of $N$ bodies can be mimicked by noise with amplitude 
${\eta}{\;}{\propto}{\;}\ln N/N$. As noted in Paper I, frozen-$N$ orbits and
smooth characteristics do indeed tend to diverge linearly on a time scale
$t_{G}{\;}{\propto}{\;}N^{1/2}t_{D}$; and, as will be seen below, this linear
divergence is well reproduced by an appropriately defined white noise with
amplitude ${\eta}{\;}{\propto}{\;}1/N$. (Given the limited dynamical range
in particle number for the simulations described in this paper -- 
$10^{2.5}{\;}{\le}{\;}N{\;}{\le}{\;}10^{5.5}$ -- it would seem impossible to
distinguish unambiguously between scalings ${\eta}{\;}{\propto}{\;}1/N$ and 
${\eta}{\;}{\propto}{\;}\ln N/N$. The simulations are consistent with both.)

When subjected to friction and white noise, chaotic orbits behave very 
differently. Comparatively weak perturbations typically induce an initial
exponential divergence from the unperturbed orbit at a rate ${\Lambda}$ that 
is comparable to the largest (short time) Lyapunov exponent ${\chi}_{S}$ for 
the unperturbed orbits\cite{HKM}. For stronger perturbations, the separation 
between perturbed and unperturbed orbits quickly becomes `macroscopic,' at 
which point the divergence become slower than exponential, albeit still more 
rapid than what is observed for regular orbits\cite{KN}. One might, therefore, 
expect that, when acting on chaotic initial conditions, discreteness effects 
associated with a finite-$N$ system would
induce (1) an initial exponential divergence at a rate ${\Lambda}$ that is
comparable to ${\chi}_{S}$, a number typically much smaller than the `true'
Lyapunov exponent ${\chi}$ associated with orbits in the frozen-$N$ system,
followed by (2) a slower subexponential divergence which is still faster
than the divergence associated with regular orbits. 

The simulations summarised
in this paper provided unambiguous confirmation of the second of these
expectations. Viewed macroscopically, frozen-$N$ orbits corresponding in the 
continuum limit to chaotic orbits typically diverge {\em linearly} from
smooth orbits with the same initial condition; but for large $N$ the time 
scale $t_{G}{\;}{\propto}\;(\ln N)t_{D}$ associated with this divergence is 
much shorter than the time scale $t_{G}{\;}{\propto}{\;}N^{1/2}t_{D}$ 
associated with regular orbits. 

Unperturbed orbits evolved in a smooth integrable potential are multiply
periodic and, as such, are characterised by Fourier spectra in which power
is concentrated at a countable set of discrete frequencies. Friction and
noise destroy this exact periodicity, resulting in a more
complex Fourier spectrum. To the extent that the friction and noise are weak, 
the orbit should remain nearly regular and the spectrum should remain `similar
to' the spectrum associated with the unperturbed orbit. However, when the
friction and noise become larger in amplitude the orbit should become `less
nearly periodic,' and the spectrum should become `more complex' than the 
spectrum associated with the unperturbed orbit. In the same sense, one might 
expect that, as $N$ decreases, frozen-$N$ orbits corresponding to regular 
orbits in a smooth potential will become `less regular' and be characterised 
by Fourier spectra that are `more complex'. As described in Paper I, this 
intuition can be made precise by computing from an orbital time series such 
quantifiable measures of 
orbital complexity\cite{KEB} as (1) the number of frequencies in a discrete 
Fourier series that contain more than some fixed fraction $j$ of the power in 
the peak frequency or (2) the minimum number of frequencies required to 
capture a fixed fraction $k$ of the total power.

By contrast, chaotic orbits in a smooth potential are in general 
aperiodic\cite{Tab}. This implies that, even in a discrete time series, their 
power should be spread over a larger number of frequencies, so that the orbit
will have substantially larger `complexity'\cite{KEB}. 
For very small $N$, where the qualitative character of the orbits is dominated
by close encounters and distinctions between chaotic and regular motions
are difficult to identify, one might expect that frozen-$N$ orbits 
corresponding in the continuum limit to regular and chaotic orbits would 
have comparable complexities, much larger than the typical complexity 
associated with a smooth regular orbit and larger even than the complexity
associated with a smooth chaotic orbit. As $N$ increases, discreteness effects
will presumably become less important and the complexity of both the regular
and chaotic frozen-$N$ orbits will decrease. For the case of regular 
frozen-$N$ orbits, the complexity should eventually converge towards the
comparatively small value associated with a smooth regular orbit. For the
case of chaotic frozen-$N$ orbits, the complexity should instead converge
towards the substantially larger value associated with a smooth chaotic orbit.

To the extent that discreteness effects associated with a fixed number $N$ 
can be successfully modelled in terms of a suitably defined noise with
amplitude ${\eta}$, one might also expect that the typical complexity $n(N)$ 
associated with a frozen-$N$ orbit with given $N$ will be comparable to
the complexity $n({\eta})$ associated with a smooth orbit with the same 
initial condition evolved in the presence of perturbations of amplitude 
${\eta}{\;}{\propto}{\;}1/N$. As described below, both these intuitions were 
in fact confirmed.

Discreteness effects can also be explored in the context of the phase mixing
of initially localised orbit ensembles, a subject which has received 
considerable attention in both galactic astronomy 
\cite{MABK}\cite{K98}\cite{KPS} and accelerator dynamics\cite{Rami}\cite{BKK} 
since Merritt and Valluri\cite{MV} coined the term `chaotic mixing' to 
characterise the much more efficient phase mixing associated with ensembles 
of chaotic orbits. 

Localised ensembles of regular initial conditions evolved in a smooth potential
will initially diverge as a power law in time and, when viewed over much longer
time scales, exhibit a coarse-grained evolution, again proceeding as a power
law in time, towards a time-independent equilibrium state. The introduction
of friction and noise accelerates the original divergence, but that divergence
still proceeds as a power law in time. Quantifying the later time evolution
is more subtle because the perturbations allow the orbits to access phase
space regions which would otherwise be inaccessible. However, what {\em is}
clear is that, as probed by various lower order moments, the ensemble evolves
exponentially in time towards a `well-mixed' state that manifests the 
symmetries of the unperturbed potential\cite{KN}. If, {\em e.g.,} the potential
admits a reflection symmetry, ${\Phi}(-x)={\Phi}(x)$, the mean value 
${\langle}x{\rangle}$ associated with the ensemble will converge exponentially
towards zero. 

By contrast, localised ensembles of chaotic orbits evolved in a smooth 
potential initially diverge exponentially at a rate ${\Lambda}$ that is 
comparable to the value of the largest Lyapunov exponent ${\chi}_{S}$ and, 
when viewed over somewhat longer time scales, exhibit a coarse-grained 
evolution, exponential in time, towards a time-independent, or nearly 
time-independent state. The rate ${\lambda}$ associated with this subsequent 
evolution is typically much smaller than ${\Lambda}$ and is not directly 
related to ${\chi}_{S}$, although loose correlations between ${\lambda}$ and 
${\chi}_{S}$ often exist\cite{MV}\cite{KN}. Subjecting these same ensembles 
to friction and noise typically increases the rate ${\Lambda}$ associated
with the initial divergence, making the orbits behave in a fashion that is
`even more chaotic.' Over sufficiently long time scales, these perturbations 
will eventually drive the ensemble towards a thermal state with a temperature 
${\Theta}$ set by the friction and noise. However, on time scales much
shorter than the natural time scale $t_{R}$ associated with the friction 
and noise, the ensemble will again evolve towards a nearly time-independent
distribution; and, in many cases, the perturbations will increase the rate
${\lambda}$ associated with this convergence towards a 
near-equilibrium\cite{KPS}\cite{KN}. 

The simulations described in this paper demonstrate that, both qualitatively
and quantitatively, discreteness effects impact phase mixing for both regular
and chaotic orbits in much the same way as these perturbations.

Perhaps the most important conclusion of the work described here is the 
apparent need to distinguish between two distinct `types' of chaos. 
Short-range 
{\em microscopic chaos} associated with close encounters between individual
masses, is a ubiquitous phenomenon for the $N$-body problem, which appears to 
be present irrespective of the bulk properties of the density distribution. 
In addition, however, there is the possibility of {\em macroscopic chaos}, 
easily identified in the continuum limit, which has specific predictable 
implications for motions in the $N$-body problem.

Section II describes the potentials that were considered and the numerical
experiments that were performed. Section III described the results derived
for individual frozen-$N$ trajectories, determining how these results scale
with $N$ and demonstrating the extent to which they can be mimicked by a
suitably defined white noise. Section IV then describes the results derived
for the phase mixing of orbit ensembles, again considering both how things
scale with $N$ and the degree to which discreteness effects can be mimicked
by noise. Section V focuses on the possibility of transitions between
`regular' and `chaotic' behaviour, a phenomenon which, especially for small
$N$, can be important for systems where the smooth potential allows a 
coexistence of large measures of both regular and chaotic orbits. 
Section VI concludes by summarising the principal conclusions and 
speculating on potential implications.

\section{DESCRIPTION OF THE NUMERICAL EXPERIMENTS}
The numerical experiments described in this paper focused on the behaviour
of orbits and orbit ensembles evolved in frozen-$N$ realisations of four
different time-independent density distributions. In the continuum limit,
two of these correspond to integrable potentials; the other two correspond
to potentials which admit
large measures of chaos. These orbits and orbit ensembles were also compared
with orbits with the same initial conditions evolved in the smooth potentials 
associated with the smooth density distributions, both with and without noise. 
Each system was normalised to have mass $M=1$, and units were so chosen that 
the gravitational constant $G=1$. The four density distributions were the 
following:
\par\noindent
1. A spherically symmetric Plummer sphere, for which 
\begin{equation}
{\rho}_{P}(r)=\left({3M\over 4{\pi}b^{3}}\right)
\left(1 + {r^{2}\over b^{2}}\right)^{-5/2}.
\end{equation}
This corresponds via Poisson's equation to a potential of the form
\begin{equation}
{\Phi}_{P}(r)=-{GM\over \sqrt{r^{2}+b^{2}}}.
\end{equation}
This integrable system was considered extensively in Paper I. As in that paper,
units were chosen such that $b=1$. 
\par\noindent
2. A constant density triaxial ellipsoid, for which 
\begin{equation}
{\rho}_{E}({\bf r})={3M\over 4{\pi}abc} \times
\cases {  m^{2} & if $m^{2}{\;}{\le}{\;}1$, \cr
               0     &  if $m^{2} > 1$, \cr}
\end{equation}
with
\begin{equation}
m^{2}=\left( {x^{2}\over a^{2}} + 
{y^{2}\over b^{2}} + {z^{2}\over c^{2}}\right).
\end{equation}
For $m{\;}{\le}{\;}1$, this yields a potential of the form
\begin{equation}
{\Phi}_{E}({\bf r})={\Phi}_{0}+{1\over 2}\left(
{\omega}_{a}^{2}x^{2}+ {\omega}_{b}^{2}y^{2} + {\omega}_{c}^{2}z^{2} \right)
\end{equation}
where the frequencies ${\omega}_{a}$, ${\omega}_{b}$, ${\omega}_{c}$, are 
related to the axis values $a$, $b$, $c$, in terms of incomplete elliptic 
integrals\cite{Ber}.
Attention focused primarily on the specific parameter values 
$a=1.95$, $b=1.50$, and $c=1.05$, which imply 
${\Phi}_{0}{\;}{\approx}{\;}-2.91758$,
${\omega}_{a}{\;}{\approx}{\;}0.9759$,
${\omega}_{b}{\;}{\approx}{\;}0.8559$, and 
${\omega}_{c}{\;}{\approx}{\;}0.7161$.
\par\noindent
3. A constant density ellipsoid perturbed by a supermassive black hole, this 
corresponding to a potential of the form 
\begin{equation}
{\Phi}_{BH}={\Phi}_{E}-{GM_{BH}\over \sqrt{r^{2}+{\epsilon}^{2}}},
\end{equation}
with ${\epsilon}=10^{-3}$.
Attention here focused on a black hole mass 
$M_{BH}=10^{-1.5}M{\;}{\approx}{\;}0.0316228M$, which yields\cite{KS2} 
a potential for which, for orbits restricted energetically to
$m{\;}{\le}{\;}1$, the phase space is largely chaotic.
\par\noindent
4. A triaxial generalisation of the Dehnen\cite{Deh} potential,
for which 
\begin{equation}
{\rho}_{D}({\bf r})={M(3-{\gamma})\over 4{\pi}abc}m^{-{\gamma}}
\left( 1+m\right)^{-(4-{\gamma})}.
\end{equation}
Attention here focused on the parameter values ${\gamma}=1$ and
$a=1.0$, $b=\sqrt{5.0/8.0}$, and $c=0.5$, values first considered by Merritt
and Fridman\cite{MF} as a prototype for a cuspy triaxial galaxy. The phase
space associated with this potential has been studied extensively\cite{MF}
\cite{Siop}
and it is known that, especially for low energies, there exist large measures
of chaotic orbits.
The smooth potential ${\Phi}_{D}$ associated with ${\rho}_{D}$ cannot be 
written analytically, so that integrations in ${\Phi}_{D}$ were performed 
using a 32nd order Gauss-Legendre integration scheme written by C. 
Siopis as part of his Ph.~D. dissertation\cite{PhD}.

Independent of energy, the characteristical orbital time scale for motion in
the constant density ellipsoid, with or without a black hole, corresponds to
a dynamical time $t_{D}{\;}{\sim}{\;}10$. (The quantity 
$1/\sqrt{G{\rho}}{\;}{\approx}{\;}3.6.$) The experiments with the Plummer
potential described in this paper were performed for intermediate energies
for which, again, $t_{D}{\;}{\sim}{\;}10$. The Dehnen potential exhibits a
much larger degree of central concentration and, as such, $t_{D}$ exhibits 
greater variability. 

The nonintegrable potential (2.6) is comparatively simple in the sense that,
for the energies and black hole masses considered here, almost all smooth
orbits are chaotic. This implies that, when considering $N$-body realisations,
one need not be much concerned about the possibility of graininess 
converting an 
initially `chaotic' orbit into a `regular' orbit. By contrast, for the 
comparatively low energies considered here the smooth Dehnen potential admits 
large measures of both regular and chaotic orbits, so that graininess could
well induce numerous transitions between `regular' and `chaotic' behaviour.

Frozen-$N$ density distributions of the form
\begin{equation}
{\rho}_{N}={1\over N}\sum_{i=1}^{N} {\delta}_{D}({\bf r}-{\bf r}_{i})
\end{equation}
were generated by randomly sampling the smooth density distributions ${\rho}$.
These correspond to $N$-body potentials 
\begin{equation}
{\Phi}_{N}({\bf r})=-{1\over N}\sum_{i=1}^{N}
{1\over \sqrt{({\bf r}-{\bf r}_{i})^{2}+{\epsilon}^{2}}}
\end{equation}
which incorporate a tiny `softening parameter' with value 
${\epsilon}{\;}{\le}{\;}10^{-3}$. 

Orbits were integrated in frozen-$N$ realisations with 
$10^{2.5}{\;}{\le}{\;}N{\;}{\le}{\;}10^{5.5}$. The integrations were 
performed with a variable time step integration scheme which was guaranteed 
to conserve the energy of each particle to at least one part in $10^{3}$. The 
energy of a typical orbit was conserved to within a few parts in $10^{5}$. 
Estimates of the largest (short time) Lyapunov exponents orbits were obtained 
in the usual way by simultaneously tracking the evolution of a small initial 
perturbation, periodically renormalised at fixed intervals 
${\Delta}t$\cite{LL}. The short time Lyapunov exponents derived in this 
fashion typically exhibited rapid convergence towards near-constant values -- 
much more rapid convergence than has been observed\cite{KS2}\cite{Siop} for 
the smooth potentials -- so that a time ${\sim}{\;}100$ dynamical, or crossing,
times $t_{D}$ was sufficient to yield reasonable estimates. 

Some of the orbital data were Fourier analysed to determine their `orbital 
complexity', defined as in \cite{KEB} as the number of frequencies required
to capture a fixed fraction $k$ of the total power. More precisely, this
entailed determining for each orbit the quantities $n_{x}$, $n_{y}$, and 
$n_{z}$, defined, respectively, as the minimum number of frequencies required 
to capture a fixed fraction $k$ of the power in each direction, and then 
assigning a total complexity
\begin{equation}
n=n_{x}+n_{y}+n_{z}.
\end{equation}
In order to obtain a reasonably sharp Fourier spectrum, orbital data were
typically recorded at intervals ${\delta}t=0.01t_{D}$ or less, and each orbit
was represented by a time series containing at least $4096$ points. 

Phase mixing in frozen-$N$ systems was explored by performing integrations
of orbit ensembles comprised (mostly) of $1600$ initial conditions localised
within a phase space hypercube of size ${\sim}{\;}10^{-3}$ the size of the
accessible phase space region. For each cell of initial conditions, 
experiments were typically repeated for several different frozen-$N$ density
distributions, each sampling the same smooth ${\rho}$, so as to extract 
improved statistics. The experiments with
$N=10^{2.5}$, $10^{3}$ and $10^{3.5}$ were each performed for six different 
frozen-$N$ density distributions, those with $N=10^{4}$ and $10^{4.5}$ for 
four distributions, and those for $N=10^{5}$ for two distributions. Because 
of computational constraints -- each orbit with $N=10^{5.5}$ averaged roughly 
three hours on a Pentium 200 workstation --, the long time ensemble 
integrations for $N=10^{5.5}$ were performed for only one ensemble, comprised 
of $800$, rather than $1600$, orbits. However, the first fifth of each
integration, in many respects the most interesting, {\em was} repeated for
a second frozen-$N$ distribution.

To test the physical intuition that discreteness effects can be mimicked by
friction and white noise in the context of a Fokker-Planck description,
orbit ensembles were also evolved in the smooth potential in the presence
of a suitably defined `white noise.' Ordinarily, discreteness effects are
modeled by considering a Langevin equation of the form\cite{van}
\begin{equation}
{d^{2}r_{a}\over dt^{2}}=-{\nabla}_{a}{\Phi}-{\eta}{dr_{a}\over dt}+F_{a},
\qquad a=x,y,z.
\end{equation}
Here ${\eta}dr_{a}/dt$ represents a dynamical friction and $F_{a}$ 
represents Gaussian white noise, which is characterised completely by its 
first two moments:
\begin{displaymath}
{\langle}F_{a}(t){\rangle}=0, \qquad (a,b=x,y,z)
\end{displaymath}
and
\begin{equation}
{\langle}F_{a}(t_{1})F_{b}(t_{2}){\rangle}=2{\eta}{\Theta}{\delta}_{ab}
{\delta}_{D}(t_{1}-t_{2}),
\end{equation}
where $D{\;}{\equiv}{\;}2{\eta}{\Theta}$ represents the diffusion constant 
entering into a Fokker-Planck description. Choosing ${\Theta}$ to equal the
initial energy then ensures that the average energy of the orbits remains
unchanged. 

This equation
is unsatisfactory here. Energy is conserved absolutely for frozen-$N$ 
orbits, so that one must also impose energy conservation on any scheme which
aims to mimic its effects. (For very small ${\eta}$, energy remains almost
conserved for very long times. However, comparatively small $N$ should 
correspond
to relatively large ${\eta}$, which implies large changes in energy and,
as such, significant changes in the phase space regions accessible to the
noisy orbit.) For this reason, the noisy integrations described here were
performed using a modified energy-conserving noise. 

What this entailed was
(1) eliminating the dynamical friction altogether, (2) again imparting random 
kicks as in eq. (2.12), but (3) renormalising the modified velocity at
each time step by an overall factor, {\em i.e.,} ${\bf v}(t+{\delta}t)\to
{\alpha}{\bf v}(t+{\delta}t)$, with ${\alpha}$ so chosen that 
$E(t+{\delta}t)=E(t)$. Modulo 
this additional complication, the noise was integrated
using a standard algorithm\cite{GSH} based on a fourth order Runge-Kutta 
integration scheme with fixed time step ${\delta}t$. Most of the integrations 
were performed for ${\delta}t=2\times 10^{-4}$, it having been confirmed that
the statistical effects of the noise were insensitive to the precise value
of ${\delta}t$ for ${\delta}t<10^{-3}$.

\section{`REGULAR' and `CHAOTIC' ORBITS}
\subsection{Ordinary Lyapunov exponents a useless diagnostic for 
macroscopic chaos}
In terms of a computation of Lyapunov exponents for individual orbits, it is 
difficult, if not impossible, to distinguish between  frozen-$N$ trajectories 
corresponding to regular orbits in a smooth potential and frozen-$N$ 
trajectories corresponding to smooth chaotic orbits. `Regular' and `chaotic' 
frozen-$N$ orbits typically have positive Lyapunov exponents which are 
comparable in magnitude, and that magnitude is typically much larger than
the magnitude of the largest Lyapunov exponent for chaotic orbits evolved in
the smooth potential.

This is illustrated by FIG.~1, the two panels of which exhibit estimates
of the largest Lyapunov exponents for frozen-$N$ realisations of the 
homogeneous ellipsoid -- both with and without a central mass -- and of the
Dehnen mass distribution. For both distributions, estimates of ${\chi}$ were 
computed for values of $N$ ranging between $N=10^{3.5}$ and $N=10^{5.5}$. 
Each curve in this FIGURE represents a mean value extracted by averaging over
four different initial conditions.

The three curves in the left panel correspond to regular initial conditions
evolved in an ellipsoid with $M_{BH}=0$ (solid line) and `sticky 
chaotic'\cite{Con} (dashed
lines) and  `wildly chaotic' (dot-dashed lines) initial conditions evolved in 
an ellipsoid with $M_{BH}=10^{-1.5}$. Three points in this panel are apparent:
(1) The estimates of ${\chi}$ computed for these three different sets of 
initial conditions are very nearly equal in magnitude. (2) Consistent with
the results described in Paper I, one observes no systematic dependence
on $N$. (3) The typical value of ${\chi}$ is 
comparatively large, much larger than the typical values ${\chi}_{S}$ 
associated with motion in the smooth potential. For the values of energy $E$
and mass $M_{BH}$ used to generate these ${\chi}$'s, `wildly chaotic' orbit
segments in the smooth potential typically have 
${\chi}_{S}{\;}{\sim}{\;}0.055$ and `sticky' segments have 
${\chi}_{S}{\;}{\sim}{\;}0.022$.

The facts that `regular' orbits with $M_{BH}=0$ and `chaotic' orbits with
$M_{BH}>0$ have ${\chi}$'s that are approximately equal, and that this
value is much larger than ${\chi}_{S}$, suggests strongly that these 
exponents reflect almost completely the effects of `microscopic' chaos 
associated with close encounters, and that the form of the bulk potential is 
largely immaterial. In both cases the frozen-$N$ orbits are moving through an
ellipsoid with the same constant density; and, since the black hole mass
$M_{BH}=10^{-1.5}{\;}{\ll}{\;}1$, the presence or absence of the black hole
does not significantly impact the natural orbital time scale.

The second panel in FIG.~1, generated for comparatively low energies orbits
evolved in the Dehnen potential,\cite{Sh2} shows an even closer agreement 
between
initial conditions corresponding to regular and chaotic orbits. However this
agreement does not directly corroborate the results derived from the first
panel. The smooth Dehnen potential admits a coexistence of large measures of 
both regular and chaotic orbits and, as will be discussed in Section V, for 
particle number $N{\;}{\le}{\;}10^{5}$, discreteness effects can trigger 
transitions between `regular' and `chaotic' behaviour on time scales much 
shorter than the integration times used to obtain estimates of the frozen-$N$
${\chi}$. Over the time scales of interest it is impossible to make clean
distinctions between `regular' and `chaotic' orbits in frozen-$N$ realisations 
of the Dehnen distribution. At any rate, the frozen-$N$ Lyapunov exponents are 
all much larger than the typical value ${\chi}_{S}{\;}{\sim}{\;}0.1$ for 
chaotic orbits evolved in the smooth Dehnen potential. 

\subsection{The divergence of $N$-body orbits from smooth characteristics}

One obvious way in which to effect macroscopic comparisons of frozen-$N$ 
orbits and smooth characteristics generated from the same initial condition 
is by computing such diagnostics as the configuration and velocity space
separations,
\begin{eqnarray}
Dr(t)&\;{\equiv}&\;|{\bf r}_{S}(t)-{\bf r}_{N}(t)| \qquad {\rm and} 
\nonumber \\
Dv(t)&\;{\equiv}&\;|{\bf v}_{S}(t)-{\bf v}_{N}(t)|,
\end{eqnarray}
where $({\bf r}_{S},{\bf v}_{S})$ and $({\bf r}_{N},{\bf v}_{N})$ denote, 
respectively, phase space coordinates for orbits in the smooth and frozen-$N$ 
density distributions. 

It was found in Paper I that, for individual regular orbits\cite{mod},
$Dr$ and $Dv$ typically grow linearly in time. A corresponding analysis for
individual chaotic orbits is somewhat less conclusive since, apparently,
different chaotic orbits can exhibit a larger degree of variety in their
behaviour. However, by averaging over an ensemble of $m$ different initial 
conditions, and computing the mean separation 
\begin{equation}
{\cal D}r(t){\;}{\equiv}{\;}{1\over n}\sum_{i=1}^{m} Dr_{i},
\end{equation}
and an analogous ${\cal D}v(t)$, one can again extract an unambiguous trend. 
As for the case of regular orbits, so also for chaotic orbits, viewed
macroscopically {\em the 
quantities ${\cal D}r$ and ${\cal D}v$ diverge linearly in time}. This is, 
{\em e.g.,} illustrated in the first seven panels of FIG.~2, which was 
generated by effecting pointwise comparisons of $800$ smooth and frozen-$N$ 
orbits in the potential (2.6) for
different values of $N$ ranging between $N=10^{2.5}$ and $N=10^{5.5}$. 
The initial conditions, chosen identically for each value of $N$, were 
selected so as to correspond to `wildly chaotic' orbits, for which a typical
short time Lyapunov exponent ${\chi}_{S}{\;}{\approx}{\;}0.055$. 

`Regular' and `chaotic' frozen-$N$ orbits are similar in the sense that they
both diverge from a smooth characteristic with the same initial condition 
as a power law in time: ${\cal D}r=A_{G}t$. However, the characteristic time 
scales involved are very different. For regular orbits in the Plummer 
potential, it was found ({\em cf.} eq. (4.8) in Paper I) that 
\begin{equation}
t_{G}{\;}{\approx}{\;}A_{G,reg}N^{1/2}t_{D}, \qquad {\rm regular}
\end{equation}
where $t_{D}$ a characteristic crossing time and $A_{G,reg}$ is a constant of 
order unity. For chaotic orbits in the Dehnen potential and the ellipsoid
plus black hole potential (2.6), one discovers instead that 
\begin{equation}
t_{G}{\;}{\approx}{\;}A_{G,cha}(\ln N) t_{D}, \qquad {\rm chaotic}
\end{equation}
where, again, $A_{G,cha}$ is of order unity.

The goodness of fit to such a logarithmic dependence is exhibited in the final
panel of FIG.~2, which exhibits growth times $t_{G}$ derived from least
squares fits to the data exhibited in the preceding panels. This panel should
be contrasted with FIG.~8 in Paper I, which exhibits an analogous curve
derived for regular orbits in the Plummer potential.

\subsection{Distinctions based on orbital `complexity'}

The fact that, as $N$ increases, frozen-$N$ orbits remain close to smooth
characteristics with the same initial condition for progressively longer 
times implies that, in terms of visual appearance, they also tend to more
closely resemble those smooth characteristics. This visual impression is 
easily corroborated by an examination of the complexity of the Fourier 
spectra, which characterise the extent to which the orbits are, or are not,
nearly periodic. For comparatively small $N$, frozen-$N$ orbits corresponding
to both regular and chaotic characteristic both look strongly aperiodic and
`wildly chaotic' in appearance. Not surprisingly, therefore, their complexities
are very large, large compared even with the complexities associated with
ordinary `wildly chaotic' orbits evolved in the corresponding smooth potential.
However, as $N$ increases, the complexities of regular and chaotic frozen-$N$ 
orbits both decrease and, for sufficiently large $N$, the complexities $n(N)$
converge towards the values $n_{S}$ associated with orbits in the smooth
potential. For frozen-$N$ orbits corresponding to regular characteristics,
this $n_{S}$ is typically quite small; for orbits corresponding to chaotic
characteristics, $n_{S}$ is typically much larger. 

Analogous results obtain for orbits evolved in the presence of ordinary
friction and noise\cite{SEK} and, as such, it would seem natural to ask
whether the observed variations in complexity resulting from changes in 
$N$ can also be mimicked by the energy-conserving noise considered in this 
paper. Overall, the answer to this would seem to be: yes.

FIG.~3 exhibits the mean complexity $n$ for representative samples of orbits
evolved in the potentials (2.2), (2.5), and (2.6). In each case, the
complexities were computed from an orbital time series that sampled an orbit 
of duration $T=512$ at intervals ${\Delta}t=0.05$. The first two panels were
generated for the Plummer and homogeneous ellipsoid potentials, 
both corresponding in the continuum limit to completely integrable motion.
The last two panels exhibit complexities computed for initial conditions 
corresponding to wildly chaotic and `sticky' chaotic orbits in the ellipsoid
plus black hole potential. In each panel, the diamonds reflect complexities
computed for frozen-$N$ orbits with different values of $N$ and the horizontal
dashed line corresponds to the mean complexity for unperturbed orbits with
the same initial conditions evolved in the corresponding smooth potential. 
The triangles reflect complexities computed for motion in the smooth potential 
perturbed by noise with ${\Theta}=1.0$ and a coefficient ${\eta}$ related to 
$N$ via a relation $\log_{10}{\eta}={\cal A}-\log_{10}N$. 

The constants ${\cal A}$ were {\em not} determined by using a least squares 
algorithm to make the two curves coincide, at least approximately. Rather, as 
described more carefully 
in Section IV, the connection between ${\eta}$ and $N$ was effected by 
demanding that, for the case of regular orbits, noise of amplitude ${\eta}$ 
and discreteness effects associated with a system of $N$ particles yield 
linear phase mixing at the same rates. For the Plummer potential, this  
requires ${\cal A}_{Plum}{\;}{\approx}{\;}0.0$. 
For the homogeneous ellipsoid, ${\cal A}_{ellip}{\;}{\approx}{\;}0.5$.
Because the black hole mass in the potential (2.6) is much smaller than the
total mass of the system -- $M_{BH}=10^{-1.5}M$ -- it should have only a
minimal effect on the characteristic orbital time scales; and, for this 
reason, the last two panels involved fits assuming ${\cal A}={\cal A}_{ellip}$.

Overall, the correspondence between frozen-$N$ and noisy complexities is
quite good, except for very small $N$ and very large ${\eta}$ where systematic
discrepancies can be seen to occur. What this means is that if the amplitude 
of the noise is tuned so as to reproduce the expected macroscopic behaviour 
associated with regular phase mixing of orbit ensembles, individual noisy 
orbits - both regular and chaotic -- will also manifest a degree of 
complexity comparable to that manifested by individual frozen-$N$ orbits.

\section{PHASE MIXING OF ORBIT ENSEMBLES}

\subsection{Divergence of initially localised ensembles}

For particle number $N<10^{3}$ or so, initially localised `regular' and 
`chaotic' ensembles both phase mix extremely rapidly in a fashion that makes 
it virtually impossible to distinguish between them. Inded, it would appear 
that, for such small $N$, mixing is dominated by discreteness effects and 
the form of the bulk potential is comparatively unimportant. However, for 
$N>10^{3.5}$ or so, it becomes possible to make clean distinctions between
the phase mixing of ensembles which correspond in the continuum limit to
regular vis-a-vis chaotic orbits. It is, for example, evident that `regular'
ensembles disperse as a power law in time, whereas chaotic ensembles disperse
in a fashion that is roughly exponential.

Consider, {\em e.g.,} the configuration dispersion ${\sigma}_{r}$ associated
with an ensemble, which satisfies
\begin{equation}
{\sigma}_{r}^{2}={\langle}r^{2}{\rangle}-{\langle}r{\rangle}^{2},
\end{equation}
with
\begin{equation}
{\langle}r^{p}{\rangle}={1\over m}\sum_{i=1}^{m}\;
(x_{i}^{2}+y_{i}^{2}+z_{i}^{2})^{p/2}.
\end{equation}

For the case of a homogeneous ellipsoid with $M_{BH}=0$, which corresponds
in the continuum limit to integrable orbits all oscillating with the same
natural frequencies, there is no phase mixing in the smooth potential, so
that ${\sigma}_{r}$ exhibits no systematic growth. If, however, the smooth
potential is replaced by a frozen-$N$ potential, one discovers instead that
${\sigma}_{r}$ grows as $t^{1/2}$. More precisely, the growth of the 
dispersion is well fit by a simple relation of the form
\begin{equation}
{\sigma}_{r}= (t/t_{G})^{1/2},
\end{equation}
where 
\begin{equation}
t_{G}(N)=ANt_{D}
\end{equation}
and $A$ is a constant of order unity. 

This $t^{1/2}$  behaviour is hardly 
surprising, corresponding as it does to the analytically predicted behaviour 
of an ensemble of `noisy' orbits all evolved in a harmonic potential with the 
same natural frequencies ({\em cf.}\cite{Ch}). The fact that $t_{G}$ exhibits 
a roughly linear dependence on $N$ is also expected, given the 
expectation\cite{Ch2} that discreteness effects should be manifested on a
relaxation time $t_{R}{\;}{\propto}{\;}N/\ln N$. The validity of the scaling
$t_{G}{\;}{\propto}{\;}N$ is illustrated in FIG.~7, which will be discussed
more carefully below.

That the growth in ${\sigma}_{r}$ is slower than linear and that the growth 
time increases with increasing $N$ are both evident from FIG.~4, 
which exhibits ${\sigma}_{r}$ for the same ensemble of initial conditions,
evolved in frozen-$N$ potentials with variable particle number ranging
between $N=10^{2.5}$ and $N=10^{5}$. 

For more generic models like the Plummer potential, where different integrable
orbits oscillate with different frequencies, the behaviour is somewhat
more complex: For smaller $N$ the dispersion still grows as $t^{1/2}$ on
a time scale consistent with eq. (4.4), whereas,
for larger $N$, the growth law is essentially linear. This behaviour can
be interpreted as arising from a competition between two effects. Regular
phase mixing, present even in the continuum limit, results in a
dispersion that grows linearly in time; whereas an additional `noise-induced 
phase mixing' typically induces a $t^{1/2}$ divergence between perturbed and
unperturbed orbits. For large $N$, discreteness effects have only
a small effect on the phase mixing, so that the frozen-$N$ ${\sigma}_{r}$
exhibits the same linear dependence on time as the smooth potential 
${\sigma}_{r}$. For smaller $N$ discreteness effects become more important 
and induce a larger amplitude $t^{1/2}$ growth. 

This behaviour is exhibited in FIG.~5, the six panels of which exhibit 
${\sigma}_{r}$ for evolution in the smooth Plummer potential and for 
frozen-$N$ evolution 
with $10^{2.5}{\;}{\le}{\;}N{\;}{\le}{\;}10^{5}$. It is evident that, for 
$N{\;}{\ge}{\;}10^{4}$ the overall growth is linear, at least until 
the dispersion `saturates' at a value ${\sigma}_{r}{\;}{\sim}{\;}0.4$, and
that this linear evolution is nearly indistinguishable from the behaviour
associated with the smooth evolution. The only obvious difference is that
discreteness effects tend to `fuzz out' the systematic oscillations associated 
with strictly periodic orbits in the unperturbed potential which yield the
large `spread' in values of ${\sigma}_{r}$ superimposed on the overall linear
growth. For smaller values of $N$, ${\sigma}_{r}(t)$ is 
better represented by a $t^{1/2}$ growth law; and, quite apparently, the
`spread' in ${\sigma}_{r}$ is substantially reduced. This reduction is an
obvious manifestation of the fact that, for smaller $N$, the orbits in the
ensemble exhibit significant deviations from periodicity, a fact manifested
by the increased complexities discussed in the preceding Section. 

Implicit in the preceding is the assumption that discreteness
effects really can be mimicked by noise. This was tested at two levels,
namely (i) a qualitative, visual comparison of plots of ${\sigma}_{r}(t)$ 
generated for both frozen-$N$ and noisy ensembles, and  (ii) a quantitative 
comparison of slopes associated with a ${\sigma}_{r}^{2}=t/t_{G}$ growth law.
The degree to which noise can mimic discrete effects is evident visually
from a comparison of individual panels in FIG.~5, which were 
generated from frozen-$N$  ensembles evolved in the Plummer potential, with 
the corresponding panels in FIG.~6, which were generated from the same 
ensembles of initial conditions, now evolved in the smooth potential in the 
presence of energy-conserving white 
noise. For particle number as small as $N=10^{2.5}$, the correspondence is
comparatively poor, the frozen-$N$ ${\sigma}_{r}$ growing considerably more
rapidly than the noisy ${\sigma}_{r}$. However, already for $N=10^{3}$ the
correspondence is quite reasonble and, for $N{\;}{\ge}{\;}10^{4}$, the
frozen-$N$ and noisy plots are virtually indistinguishable. 

The preceding comparison of frozen-$N$ ensembles with fixed $N$ and noisy 
ensembles evolved with ${\eta}{\;}{\propto}{\;}1/N$ was motivated by the 
fact that the noisy dispersions were also well fit by a $t^{1/2}$ growth law 
where, however,
\begin{equation}
t_{G}=Bt_{D}/{\eta},
\end{equation}
with $B$ a constant of order unity. A comparison of eqs. (4.4) and
(4.5) implies a natural identification 
\begin{equation}
\log_{10}{\eta}={\cal A}-\log_{10}N,
\end{equation}
with ${\cal A}$ yet another constant. For the ellipsoid potential, the best
fit ${\cal A}=0.0{\pm}0.1$. For the Plummer potential ${\cal A}=0.6{\pm}0.1$.
The identification between FIGS.~5 and 6 was effected assuming 
${\cal A}=0.5$. The extent to which the growth rates for frozen-$N$ and
noisy orbits can be related by the simple relation (4.6) can be gauged from 
FIGS.~7, which superimpose plots of $t_{G}(N)$ and $t_{G}({\eta})$ for the
Plummer and ellispoid potentials, with $N$ and ${\eta}$ related by eq. (4.6).

The behaviour exhibited by ensembles corresponding to chaotic orbits is very
different. In this case, the dispersion associated with an ensemble evolved
in the smooth potential typically diverges exponentially in time, and 
discreteness effects only serve to accelerate this growth. This is, {\em e.g.,}
evident from FIG.~8, which plots $\ln {\sigma}_{r}$ for an ensemble of initial
conditions evolved in the potential (2.6) that corresponds in the continuum
limit to `wildly chaotic' orbits. It is apparent that the growth of 
${\sigma}_{r}$ is not strictly exponential, but it certainly {\em is} faster
than the exponential growth associated with the smooth potential, which is
exhibited in the final panel. One other point is also evident: For the
case of the Plummer potential, a frozen-$N$ integration with $N$ as small as 
$N{\;}{\sim}{\;}10^{4.5}$ yields phase mixing almost identical to that 
observed in the smooth potential. For the case of the chaotic ensemble
evolved in the potential (2,6), 
even $N=10^{5.5}$ yielded phase mixing that was {\em much more rapid} than
that associated with the corresponding smooth potential. 

FIG.~10 exhibits an analogous plot, again generated for the potential (2.6),
but now considering an ensemble of initial conditions which, in the 
continuum limit, correspond to comparatively `sticky' chaotic orbits which
initially disperse much more slowly. (The typical value of the largest short
time Lyapunov exponent for smooth orbits in this `sticky' ensemble was
${\chi}(t=256){\;}{\sim}{\;}0.22$. The typical value for the `wildly chaotic'
ensemble was ${\chi}{\;}{\sim}{\;}0.55$.) For $N{\;}{\le}{\;}10^{4}$ or so,
FIGS.~8 and 10 are comparatively similar, the `stickiness' manifested
in the continuum limit being largely lost. However, for larger $N$ more
conspicuous differences become apparent. It is, for example, clear that, for
particle number as large as $N=10^{5.5}$, the dispersion for the `sticky'
ensemble only becomes `macroscopic' on a time scale appreciably longer 
than the time scale for the `wildly chaotic' ensemble. 

To a considerable degree, the behaviour exhibited by chaotic frozen-$N$
ensembles is again well-mimicked by energy-conserving noise. This is, 
{\em e.g.,} evident from FIGS.~9 and 11 which, respectively, exhibit
the same ensembles of initial conditions as FIGS.~8 and 10, now evolved
in the smooth potential in the presence of energy-conserving white noise. 
The first seven panels in each of these FIGURES involved the same 
identification between $N$ and ${\eta}$ as did FIG.~6, this motivated
by the recognition that the black hole mass 
$M_{BH}=10^{-1.5}$ is too small to significantly alter the natural 
dynamical time $t_{D}$. The final panel in each FIGURE exhibits 
$\ln {\sigma}_{r}$ for ${\eta}=10^{-7.5}$, the largest value of ${\eta}$ that 
does {\em not} result in phase mixing that is significantly more rapid than
that associated with motion in the unperturbed smooth potential. Presuming
that the correspondence between $N$ and ${\eta}$ established here can be 
extrapolated to larger $N$ and smaller ${\eta}$, it follows that, for the
ensembles considered here, one would require particle number as large as
$N{\;}{\sim}{\;}10^{8}$ before discreteness effects become unimportant over
the time scales of interest!

One obvious issue is the extent to which the behaviour of
orbit ensembles evolved in different frozen-$N$ realisations of the same
potential with the same number $N$ is, or is not, the same. For the case
of the integrable Plummer and ellipsoid potentials, it would appear that,
even for numbers as small as $N{\;}{\sim}{\;}10^{3}$, the statistical 
properties of orbit ensembles are essentially the same for different frozen-$N$
realisations. This, however, is {\em not} true for chaotic ensembles. In
this case, one appears to require $N>10^{5}$ or so before noticeable 
differences between
different realisations are suppressed. This is, {\em e.g.,} evident from
FIG.~12, which exhibits ${\sigma}_{r}$ for the same initial conditions
used to generate FIGS.~10 and 11, now focusing on a time interval only one
fifth as long and including with dotted lines the results of the different
realisations which were averaged to yield the solid line. 

\subsection{Approach towards a `well-mixed' state}
The fact that initially localised ensembles of chaotic orbits evolved in a
smooth potential should diverge exponentially is more or less obvious; and
the simulations described in this paper show that graininess associated with 
finite $N$ has the same qualitative effect as noise with characteristic
amplitude ${\eta}{\;}{\propto}{\;}1/N$. Frozen-$N$ evolution results in a
divergence which is even faster than that associated with the unperturbed
ensemble, albeit no longer strictly exponential. 

Less obvious, but also true, is the fact that chaotic ensembles integrated in
a smooth potential tend to evolve towards a near-equilibrium state, and 
that this evolution typically proceeds exponentially in time. For systems
manifesting the reflection symmetries associated with the potentials explored
in this paper, this implies in particular that moments like  
${\langle}x{\rangle}$ and ${\langle}v_{x}{\rangle}$ tend to zero exponentially.
The obvious question then is whether discreteness effects, which can 
significantly accelerate the initial rate at which ensembles disperse, also
act to accelerate the rate at which ensembles evolve towards a 
near-equilibrium. 

Similarly, one can consider the effects of graininess on regular orbits. 
Unperturbed regular orbits evolved in a smooth potential do {\em not} exhibit
a rapid exponential approach towards a (near-)equilibrium corresponding to a
finite phase space volume. Rather, what one observes
is a more modest power law evolution towards a near-uniform population 
of the invariant tori to which they are restricted. Allowing for discreteness
effects allows the orbits to escape from these invariant tori and, as such,
one might again ask: is there an approach towards some more general
`near-equilibrium'? And, if so, how does this approach proceed in time?

For the case of potentials which, in the continuum limit, correspond to 
integrable systems, one finds that `weak' discreteness effects associated
with comparative large $N$ have only a minimal effect on such moments as
${\langle}x{\rangle}$. If, however, one allows for smaller $N$ systems,
where discreteness effects become more important, these moments will in fact
exhibit an unambiguous exponential convergence towards zero. This is, 
{\em e.g.,} illustrated in FIGS.~13 and 14, which exhibit the quantity
$\ln |{\langle}z{\rangle}|$ for frozen-$N$ ensembles evolved, respectively, 
in the integrable ellipsoid and Plummer potentials.
The fact that this convergence seems to terminate at 
$\ln |{\langle}z{\rangle}|{\;}{\sim}{\;}-3$ is a finite sampling effect, 
reflecting the fact that the ensembles were comprised of only $1600$ orbits.
Even if one were to select `at random' 1600 points from a continuous 
distribution with ${\langle}z{\rangle}=0$, one would have a sample for which
${\langle}z{\rangle}{\;}{\ne}{\;}0$.

For the case of chaotic potentials one observes an exponential decrease in
such moments, even in the continuum limit, but discreteness effects again
serve to increase the rate associated with this evolution. This is, {\em e.g.,}
illustrated in FIGS.~15 and 17, which exhibit 
$\ln |{\langle}z{\rangle}|$ for the `wildly chaotic' and `sticky' ensembles
used to generate FIGS.~(3.5) and (3.7). The degree to which this enhanced
exponential evolution can be mimicked by energy-conserving noise is illustrated
in FIGS.~16 and 18. As for FIGS.~(3.6) and (3.8), the final panel in
each of these FIGURES corresponds to noise with amplitude ${\eta}=10^{-7.5}$
which, once again, is the weakest noise not to occasion a significant increase
in the overall efficacy of chaotic mixing.

\section{Transitions between `regular' and `chaotic' behaviour}
It is well known that friction and noise can convert a regular orbit into a
chaotic orbit and vice versa. Suppose, {\em e.g.,} that an initially chaotic
orbit is evolved in a smooth potential in the presence of noise for some 
finite period and that the noise is then turned off. If the evolution is 
continued in the absence of noise, one may then find that the orbit has become
regular with no positive Lyapunov exponents. Why this can happen is easy to
understand: Noise serves to continually `bump' the orbit from one smooth
characteristic to another and it is quite possible that such `bumps' will
eventually deflect the orbit from a chaotic to a regular characteristic. To
the extent that discreteness effects can be mimicked by noise, one would
anticipate the possibility of similar transitions in an $N$-body 
evolution\cite{Bohn}.

This was not an issue for orbits evolved in the potentials (2.2) and (2.5),
both of which are integrable. This possibility is also unimportant for the
nonintegrable black hole plus ellipsoid potential discussed above since, for
the energies and black hole masses that were considered, the energetically
accessible phase space is almost completely chaotic. However, the possibility 
of such transitions is a major issue for the Dehnen potential where, for most 
energies, the constant energy hypersurface admits large measures of both 
regular and chaotic orbits. 

One hint that such transitions might be present derives from a computation
of $Dr$, the mean separation between frozen-$N$ orbits and smooth 
characteristics with the same initial condition, generated from ensembles of
initial conditions corresponding to only regular or only chaotic orbits. 
One example is provided in FIG.~19, which exhibits $Dr(t)$ for ensembles
evolved in frozen-$N$ realisations of the Dehnen distribution with $N=10^{4}$, 
$N=10^{4.5}$, and $N=10^{5}$. The three left panels were generated from 
regular initial conditions; the three right panels from chaotic initial 
conditions. Both sets of initial conditions were selected to have very low
energies, so that they were restricted to the central portions of the 
system\cite{Sh1}. 

It is clear that the chaotic initial conditions
yield a more `irregular' time dependence although, as expected, in both cases 
$Dr$ evidences a nearly linear growth in time. However, it is also obvious 
that the growth time $t_{G}$ for the chaotic initial conditions is {\em not}
much longer than the growth time for the regular initial conditions. For all
three values of $N$, the best fit value of $t_{G}$ is comparable for the
regular and chaotic initial conditions although, for $N=10^{5}$, it is clear
that $Dr$ grows somewhat more slowly for the regular orbits. The obvious
interpretation is that, even at very early times, a significant fraction of
the initial conditions have switched between `regular' and `chaotic' behaviour.

Another, even more striking, indication that such transitions occur is
provided by computing distributions of orbital complexities. This is 
illustrated in FIG.~20, which exhibits the complexity $n$ for the low energy
ensembles
used to generate FIG.~19, as well as for a comparable set of integrations
performed for ensembles with substantially higher energies\cite{Sh2}.
The three left panels correspond to the lower energy initial conditions;
the three right panels to the higher energy initial conditions. In each
case, the chaotic orbits are represented by a solid curve and the regular
orbits by a dashed curve. 

It is clear that, for the lower energy evolutions,
the distributions of complexities for the regular and chaotic initial
conditions are almost identical, although there is some hint that, for
$N=10^{5}$, the regular orbits are somewhat `less complex.' However, for the 
higher energy evolutions regular and chaotic initial conditions yield very 
different distributions of complexity. For all three
values of $N$, the distribution for the chaotic ensembles is unimodal and
peaked at a value $n{\;}{\sim}{\;}500-600$. For the two larger values of
$N$, the distributions for the regular orbits are also singly peaked, but 
the peaks occur at significantly lower values $n{\;}{\sim}{\;}200$. 
Significantly, though, for the lowest value of $N$ the distribution of 
complexities for the regular initial conditions is distinctly bimodal, with
one peak at a value similar to that obtained for the chaotic initial conditions
and another at a significantly smaller value. This sort of bimodal distribution
is exactly what one would expect for an ensemble of orbits which is partially
regular and partially chaotic\cite{SEK}.

Direct proof that transitions between regularity and chaos occur, and an
estimate as to the time scale on which such transitions occur,
is straightforward to obtain. Given orbital data from frozen-$N$ ensembles
integrated for 
a total time $T$, one can use snapshots at earlier times ${\tau}<T$  to
generate new ensembles of initial conditions and integrate those ensembles
in the smooth potential to compute Lyapunov exponents. Comparing the results
of such integrations with integrations of the original initial conditions
in the smooth potential then permits a concrete probe of the extent to which
discreteness effects have made a regular ensemble `less regular' and/or a
chaotic ensemble `less chaotic.'\cite{st}

The results of such an analysis are summarised in FIG.~22, which plots
the mean short time Lyapunov exponent ${\langle}{\chi}_{S}{\rangle}$ for
smooth orbit ensembles as a function of ${\tau}$. The three left panels 
were generated for the low energy ensembles; the three right panels for the
high energy panels. In each case the initially chaotic ensembles are
represented by diamonds and a solid line; the initially regular ensembles
are represented by triangles and a dashed line. The three rows again correspond
respectively to $N=10^{4}$, $N=10^{4.5}$, and $N=10^{5}$. 

In each case it is clear that, overall the mean ${\chi}_{S}$ for the initially
chaotic ensembles decreases systematically with ${\tau}$ and that the mean
${\chi}_{S}$ for the initially regular ensembles increases. This is of course
exactly what would be expected if, as ${\tau}$ increases, progressively
larger numbers of transitions between regularity and chaos have occured during
the frozen-$N$ evolution. Indeed, to the extent that the two ensembles of
initial conditions sample the same phase space regions, one would expect that
the values of ${\chi}_{S}$ for the regular and chaotic ensembles should 
converge to a common value. It is evident that, for the lower energy ensemble,
convergence or near-convergence has in fact been achieved for the largest 
values of ${\tau}$, and it is also apparent that convergence happens more
rapidly for smaller $N$. This again is exactly as expected. Larger $N$ 
corresponds to weaker `noise,' but weaker noise should be less effective in
inducing transitions. 

{\em A priori} it might seem surprising that the
mean ${\chi}_{S}$ for the regular orbits approaches its limiting value much
more quickly than does the mean for the chaotic orbits. This, however, is
likely a manifestation of the phase space structure associated with the Dehnen
potential. For both low and high energies, the initially regular ensemble 
corresponds in the smooth potential to box orbits which occasionally pass
quite close to the center of the system; but, in the center, the regular
and chaotic phase space regions are entangled in a very complex fashion, so
that it is comparatively simple for initially regular orbits to be deflected
to chaotic trajectories.

\section{DISCUSSION}

The experiments described in this paper lead to several unambiguous 
conclusions regarding the behaviour of orbits and orbit ensembles evolved
in frozen-$N$ realisations of smooth density distributions.

It is, for example, clear that ordinary Lyapunov
exponents computed for individual frozen-$N$ trajectories do {\em not} provide
a useful characterisation of the degree of {\em macroscopic chaos} manifested
by these trajectories. Different initial conditions with the same energy
evolved in frozen-$N$ realisations of a specified potential typically have
Lyapunov exponents that are comparable in magnitude and exhibit little if any
dependence on $N$, even though, in terms of their bulk properties, one may
look very nearly regular and another wildly chaotic.

This can be interpreted by asserting that, at least for large $N$, one can
make comparatively clear distinctions between two types of chaos which
may be associated with the $N$-body problem. On the one hand, there is
{\em microscopic chaos} associated with close encounters between nearby
masses. This chaos, which is presumably responsible for the large Lyapunov
exponents associated with frozen-$N$ orbits, is (almost) always present but,
at least for comparatively large $N$, tends to `wash out' macroscopically.
On the other hand, there is the possibility of {\em macroscopic chaos}
which, if present in the continuum limit, will also have manifestations in
frozen-$N$ simulations.

The experiments also demonstrate a clear sense in which, as $N$ increases,
frozen-$N$ trajectories become `more similar' to smooth potential
characteristics generate from the same initial condition. In particular, this
implies that it becomes progressively easier to distinguish between `regular'
and `chaotic' macroscopic behaviour. This similarity can be quantified in at 
least three different ways: 
\par\noindent
1. {\em The rate at which frozen-$N$ orbits and smooth characteristics 
diverge.}
For both regular and chaotic initial conditions, frozen-$N$ trajectories and 
smooth characteristics tend to diverge linearly in time. However, the 
$N$-dependence of the time scale $t_{G}$ associated with this divergence 
differs dramatically for regular and chaotic orbits. For initial conditions
corresponding to regular orbits, $t_{G}{\;}{\propto}{\;}N^{1/2}$. For
initial conditions corresponding to chaotic orbits,
$t_{G}{\;}{\propto}{\;}\ln N.$ It follows that, for large $N$, frozen-$N$
trajectories and smooth characteristics corresponding to regular orbits remain
`close' much longer than do trajectories and characteristics corresponding
to chaotic orbits.
\par\noindent
2. {\em The `complexity' of Fourier spectra constructed from orbital time 
series.} For small $N$, both regular and chaotic initial conditions will, when 
integrated into the future, yield Fourier spectra that are much more complex
than the spectra associated with an evolution in the smooth potential. In
particular, both regular and chaotic orbits will yield spectra of comparable
complexity. However, as $N$ increases the complexities decrease and, for
sufficiently large $N$, one sees a convergence towards the complexities
appropriate for smooth characteristics with the same initial conditions,
be these either regular or chaotic.
\par\noindent
3. {\em The bulk properties of phase mixing, as probed, {\em e.g.} by lower 
order moments.} 
Phase mixing of initially localised ensembles in frozen-$N$ systems
is invariably more efficient than phase mixing in the corresponding smooth
potential but, as $N$ increases, the observed evolution comes to more closely
resemble phase mixing of the same initial conditions evolved in the smooth 
potential.

At least for comparatively large $N$, many discreteness effects can be well 
mimicked by energy-conserving white noise with amplitude 
${\eta}{\;}{\propto}{\;}1/N$. This is in close agreement with naive 
expectations based on the modeling of discreteness effects as a sequence of 
incoherent binary encounters, which would suggest 
${\eta}{\;}{\propto}{\;}(\ln N)/N$; and indeed, the simulations are
also consistent with the latter dependence. Modeling discreteness effects in
terms of noise works well both in predicting the expected complexities of
individual frozen-$N$ trajectories and moments associated with phase mixing.
This indicates that, at least for large $N$ {\em noise can be used to model 
both bulk statistical properties of orbit ensembles {\em and} qualitative 
properties of individual orbits.}
This suggests strongly that investigations of the effects of white noise on 
orbits in nonintegrable potentials -- which are much less expensive
computationally than frozen-$N$ integrations -- can be interpreted as 
providing information about the effects of graininess. 

Nevertheless, it is clear that modeling discreteness effects by white noise is 
{\em not} completely satisfactory. The agreement between frozen-$N$ and noisy
integrations is poor for comparatively small $N$; and noisy integrations of the
form described in this paper cannot be used to obtain estimates of the 
Lyapunov exponents associated with frozen-$N$ orbits. The development of a
`more realistic' noise and its use to model systems with $N{\;}{\gg}{\;}10^{5}$
is currently underway.

Discreteness effects can induce transitions between different orbit `types,'
including both transitions between regular and chaotic behaviour, which are
impossible in the smooth potential, and transitions between (say) `sticky'
and `wildly chaotic' behaviour which, in the smooth potential, is not 
forbidden but typically occurs only on a much longer time scale. Not 
surprisingly these transitions appear to be more common for smaller $N$; and
indeed, the simulations are consistent with the interpretation that, for
sufficiently large $N$, transitions between regular and chaos become 
essentially impossible and transitions between sticky and wildly chaotic
behaviour happen no more quickly than in the continuum limit.

Finally, and perhaps most importantly, it would appear that, at least in terms
of macroscopic properties, {\em it does make sense to speak of a smooth
$N\to\infty$ continuum limit. However, convergence towards this limit is much 
slower for density distributions which, in the continuum limit, correspond to 
nonintegrable potentials that admit chaotic orbits.}
\acknowledgments
It is a pleasure to acknowledge useful discussions with Court Bohn, Ilya 
Pogorelov, and Christos Siopis. 
Partial financial support was provided by NSF AST-0070809. The computations
involving orbital ensembles in the Dehnen potential were performed using 
resources of the National Energy Research Scientific
Computing Center, which is supported by the Office of Science of the U.S.
Department of Energy under Contract No. DE-AC03-76SF00098.
\vfill\eject

\vfill\eject
\pagestyle{empty}
\begin{figure}[t]
\centering
\centerline{
        \epsfxsize=8cm
        \epsffile{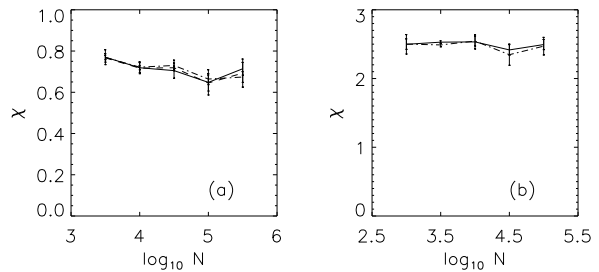}
           }
        \begin{minipage}{10cm}
        \end{minipage}
        \vskip -0.0in\hskip -0.0in
\caption{(a) Estimates of the largest Lyapunov exponent for orbits evolved
in frozen-$N$ realisations of a homogeneous ellipsoid: integrable initial
conditions evolved with $M_{BH}=0$ (solid line), `sticky' initial conditions
evolved with $M_{BH}=10^{-1.5}$ (dashed line), and `wildly chaotic' initial
conditions with $M_{BH}=10^{-1.5}$ (dot-dashed line). (b) Lyapunov exponents
evolved in frozen-$N$ realisations of the Dehnen potential for low energy
regular (solid line) and chaotic (dashed line) initial conditions.}
\end{figure}

\pagestyle{empty}
\begin{figure}[t]
\centering
\centerline{
        \epsfxsize=8cm
        \epsffile{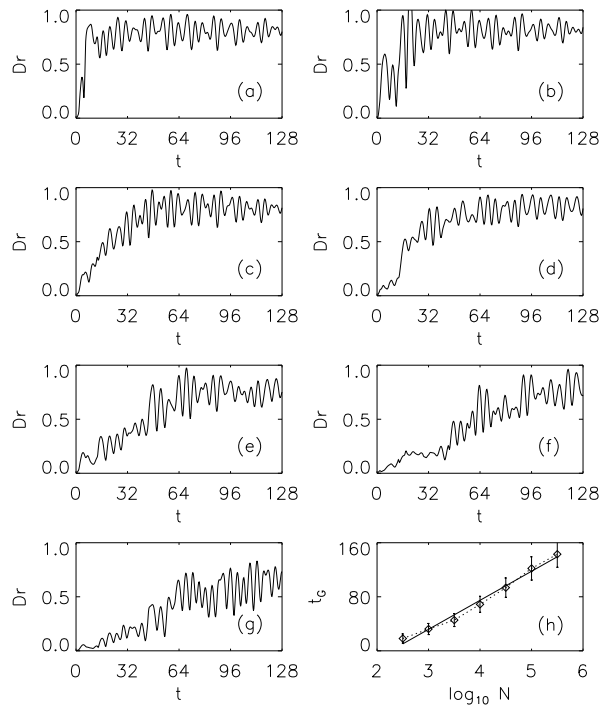}
           }
        \begin{minipage}{10cm}
        \end{minipage}
        \vskip -0.0in\hskip -0.0in
\caption{The mean separation, ${\cal D}r$, between frozen-$N$ orbits and
smooth characteristics with the same initial conditions, computed for
ensembles of $800$ chaotic initial conditions evolved in the potential
(2.6) for variable $N$. (a) $N=10^{2.5}$. (b) $N=10^{3}$. (c) $N=10^{3.5}$. 
(d) $N=10^{4}$. (e) $N=10^{4.5}$. (f) $N=10^{5}$. (g) $N=10^{5.5}$. (h) The 
growth rate $t_{G}(N)$
extracted from the preceding panels, assuming a linear growth law. The solid
line overlays a least squares fit $t_{G}=A+B\log_{10} N$.}
\end{figure}
\vfill\eject

\pagestyle{empty}
\begin{figure}[t]
\centering
\centerline{
        \epsfxsize=8cm
        \epsffile{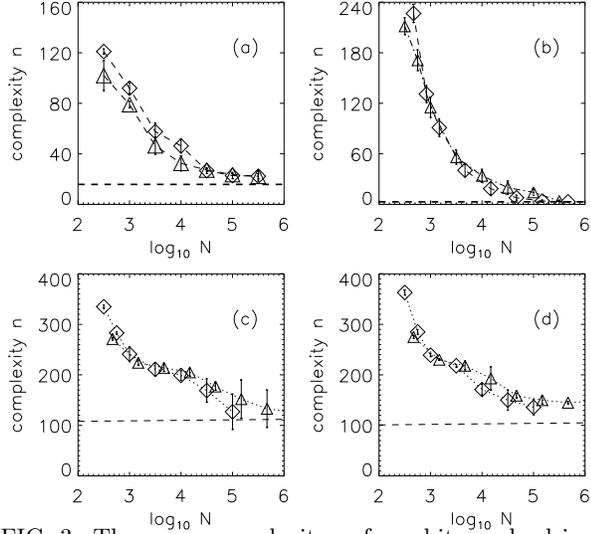}
           }
        \begin{minipage}{10cm}
        \end{minipage}
        \vskip -0.3in\hskip -0.0in
\caption{The mean complexity $n$ for orbits evolved in the potentials (2.2),
(2.5), and (2.6), considering both frozen-$N$ orbits with variable $N$
(diamonds) and smooth orbits perturbed by noise with ${\Theta}=1.0$ and
variable ${\eta}$ (triangles). ${\eta}$ was related to $N$ by the relation
${\eta}=e^{{\cal A}}/N$, with ${\cal A}$ determined as described in the text.
The dashed horizontal line exhibits the mean complexity of orbits with the
same initial conditions evolved in the smooth potential in the absence of
noise.
(a) Regular orbits in the Plummer potential. (b) Regular orbits in the
homogeneous ellipsoid potential. (c) Wildly chaotic orbits in the ellipsoid
plus black hole potential. (d) `Sticky' chaotic orbits in the ellipsoid plus
black hole potential.}
\end{figure}

\pagestyle{empty}
\begin{figure}[t]
\centering
\centerline{
        \epsfxsize=8cm
        \epsffile{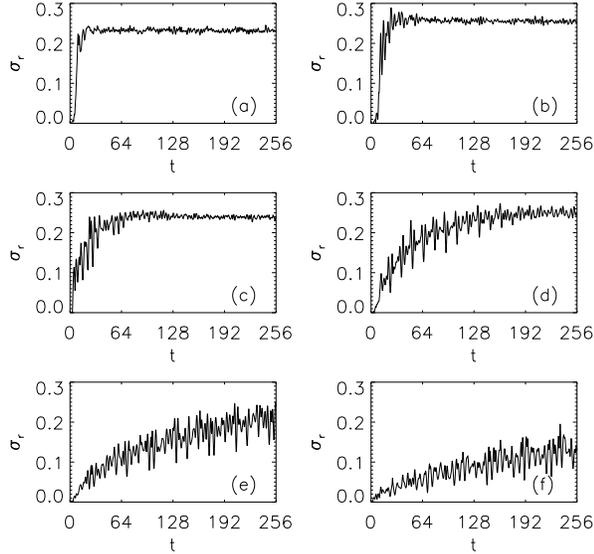}
           }
        \begin{minipage}{10cm}
        \end{minipage}
        \vskip -0.0in\hskip -0.0in
\caption{The configuration dispersion ${\sigma}_{r}$ associated 
with an
initially localised ensemble evolved in frozen-$N$ realisations of the 
integrable ellipsoid potential (2.5) for variable $N$. (a) $N=10^{2.5}$. (b) 
$N=10^{3}$. (c) $N=10^{3.5}$. (d) $N=10^{4}$. (e) $N=10^{4.5}$. (f) 
$N=10^{5}$.}
\vspace{-0.0cm}
\end{figure}
\vfill\eject
\begin{figure}[t]
\centering
\centerline{
        \epsfxsize=8cm
        \epsffile{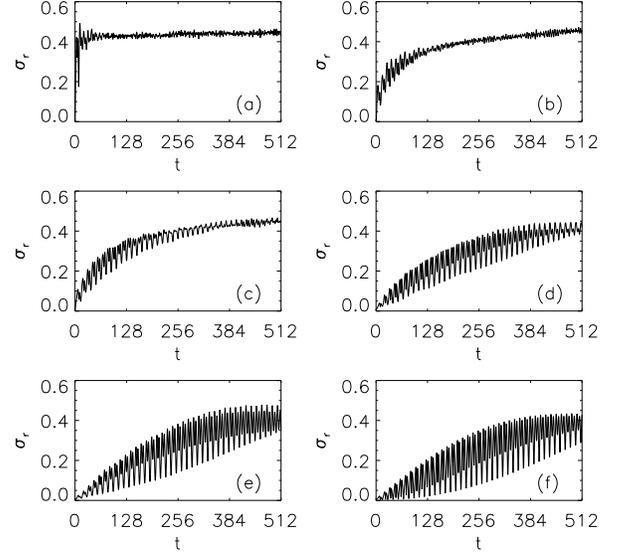}
           }
        \begin{minipage}{10cm}
        \end{minipage}
        \vskip -0.0in\hskip -0.0in
\caption{The configuration dispersion ${\sigma}_{r}$ associated with an
initially localised ensemble evolved in frozen-$N$ realisations of the 
integrable Plummer potential (2.2) for variable $N$. (a) $N=10^{3}$.
(b) $N=10^{3.5}$. (c) $N=10^{4}$. (d) $N=10^{4.5}$. (e) $N=10^{5}$. (f) 
Evolution in the smooth potential.}
\vspace{-0.0cm}
\end{figure}

\begin{figure}[t]
\centering
\centerline{
        \epsfxsize=8cm
        \epsffile{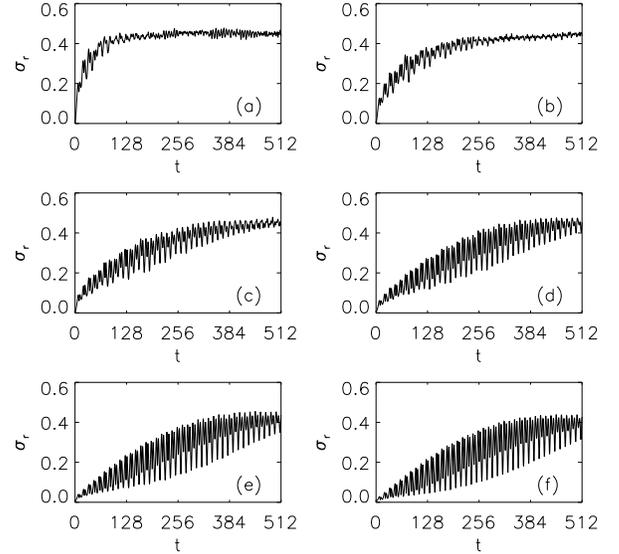}
           }
        \begin{minipage}{10cm}
        \end{minipage}
        \vskip -0.0in\hskip -0.0in
\caption{The configuration dispersion ${\sigma}_{r}$ associated with an
initially localised ensemble evolved in the smooth Plummer potential (2.2),
perturbed by noise with ${\Theta}=1.0$ and variable ${\eta}$. 
(a) ${\eta}=10^{-3}$. (b) ${\eta}=10^{-3.5}$. (c) ${\eta}=10^{-4}$.
(d) ${\eta}=10^{-4.5}$. (e) ${\eta}=10^{-5}$. (f) ${\eta}=10^{-5.5}$.}
\vspace{-0.0cm}
\end{figure}

\begin{figure}[t]
\centering
\centerline{
        \epsfxsize=8cm
        \epsffile{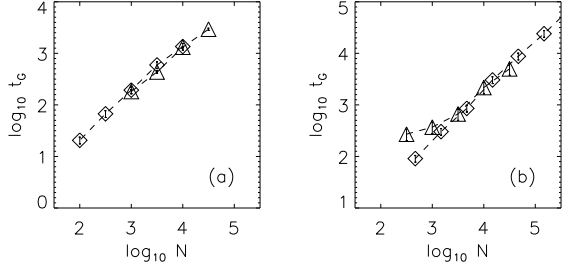}
           }
        \begin{minipage}{10cm}
        \end{minipage}
        \vskip -0.0in\hskip -0.0in
\caption{(a) The growth time $t_{G}(N)$ for the dispersion ${\sigma}_{r}$ for
orbit ensembles evolved in frozen-$N$ realisations of the Plummer potential
(diamonds) and the corresponding growth time $t_{G}({\eta})$ for orbit
ensembles evolved in the smooth Plummer potential in the presence of 
energy-conserving white noise with ${\Theta}=1.0$ and 
$\log_{10}{\eta}=-\log_{10}N$ (triangles). (b) 
$t_{G}(N)$ and $t_{G}({\eta})$ for orbit ensembles in the homogeneous 
ellipsoid potential, now setting $\log_{10}{\eta}=-\log_{10}N + 0.6$.}
\end{figure}

\begin{figure}[t]
\centering
\centerline{
        \epsfxsize=8cm
        \epsffile{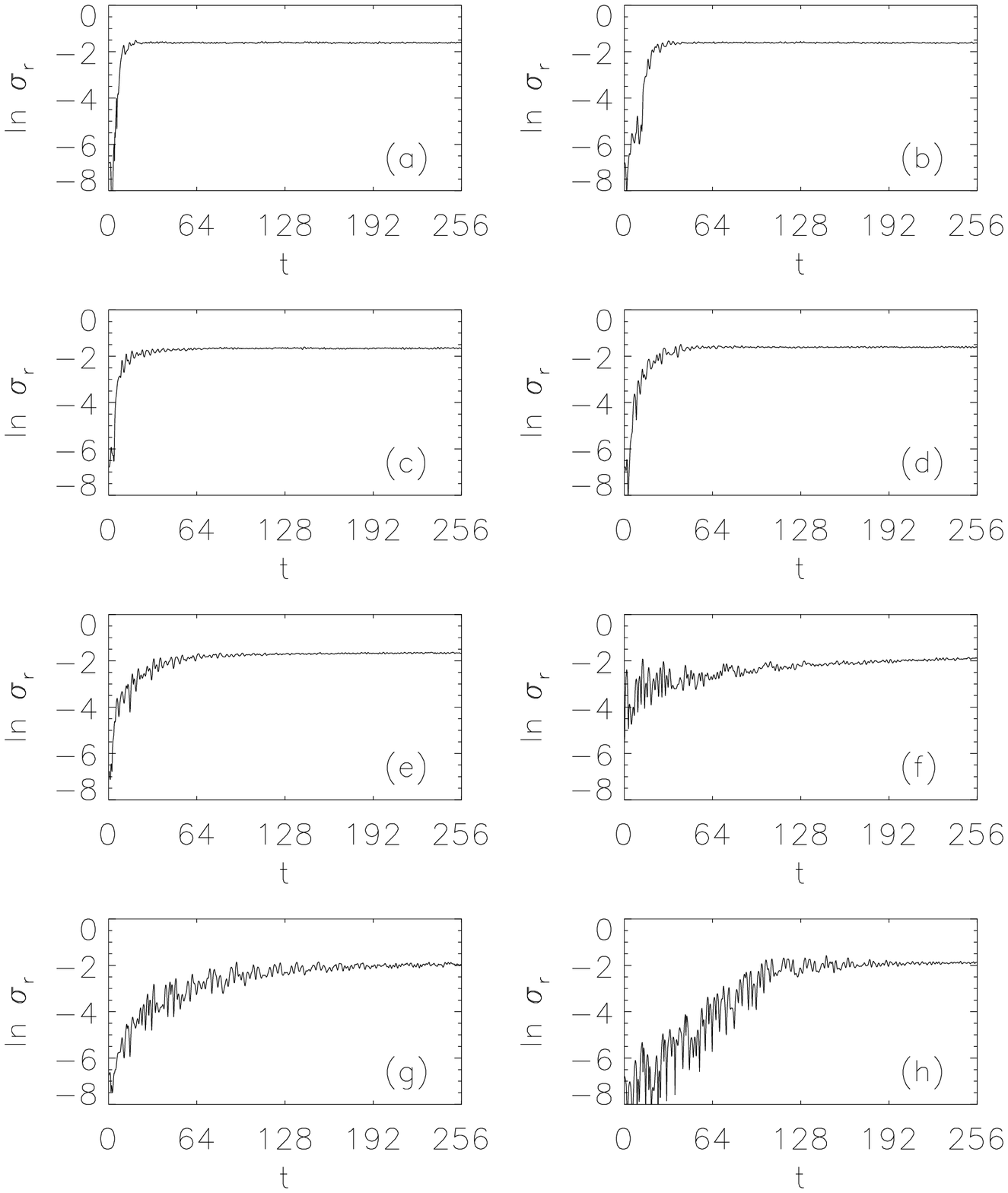}
           }
        \begin{minipage}{10cm}
        \end{minipage}
        \vskip -0.2in\hskip -0.0in
\caption{The configuration dispersion ${\sigma}_{r}$ associated with an
initially localised ensemble of wildly chaotic orbits evolved in frozen-$N$ 
realisations of the nonintegrable potential (2.6) for variable $N$. (a) 
$N=10^{2.5}$. (b) $N=10^{3}$. (c) $N=10^{3.5}$. (d) $N=10^{4}$. 
(e) $N=10^{4.5}$. (f) $N=10^{5}$. (g) $N=10^{5.5}$. (h) Evolution in the 
smooth potential.}
\vspace{-0.0cm}
\end{figure}
\vfill\eject.
\vskip 3in
\begin{figure}[t]
\centering
\centerline{
        \epsfxsize=8cm
        \epsffile{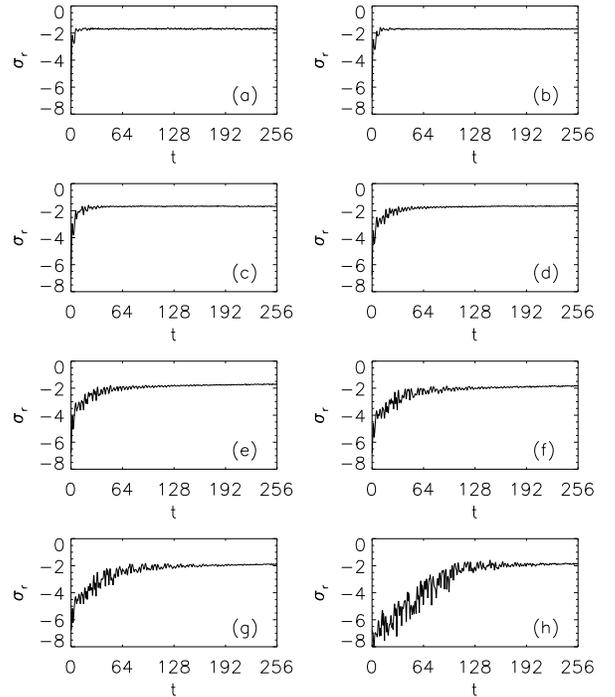}
           }
        \begin{minipage}{10cm}
        \end{minipage}
        \vskip -0.2in\hskip -0.0in
\caption{The configuration dispersion ${\sigma}_{r}$ associated with an
initially localised ensemble of wildly chaotic orbits evolved in the
nonintegrable potential (2.6) but perturbed by energy-conserving white noise 
with ${\Theta}=1.0$ and 
variable ${\eta}$. 
(a) ${\eta}=10^{-2}$. (b) ${\eta}=10^{-2.5}$. (c) ${\eta}=10^{-3}$.
(d) ${\eta}=10^{-3.5}$. (e) ${\eta}=10^{-4}$. (f) ${\eta}=10^{-4.5}$.
(g) ${\eta}=10^{-5}$, (h) ${\eta}=10^{-7.5}$.}
\vspace{-0.0cm}
\end{figure}
\vfill\eject

\begin{figure}[t]
\centering
\centerline{
        \epsfxsize=8cm
        \epsffile{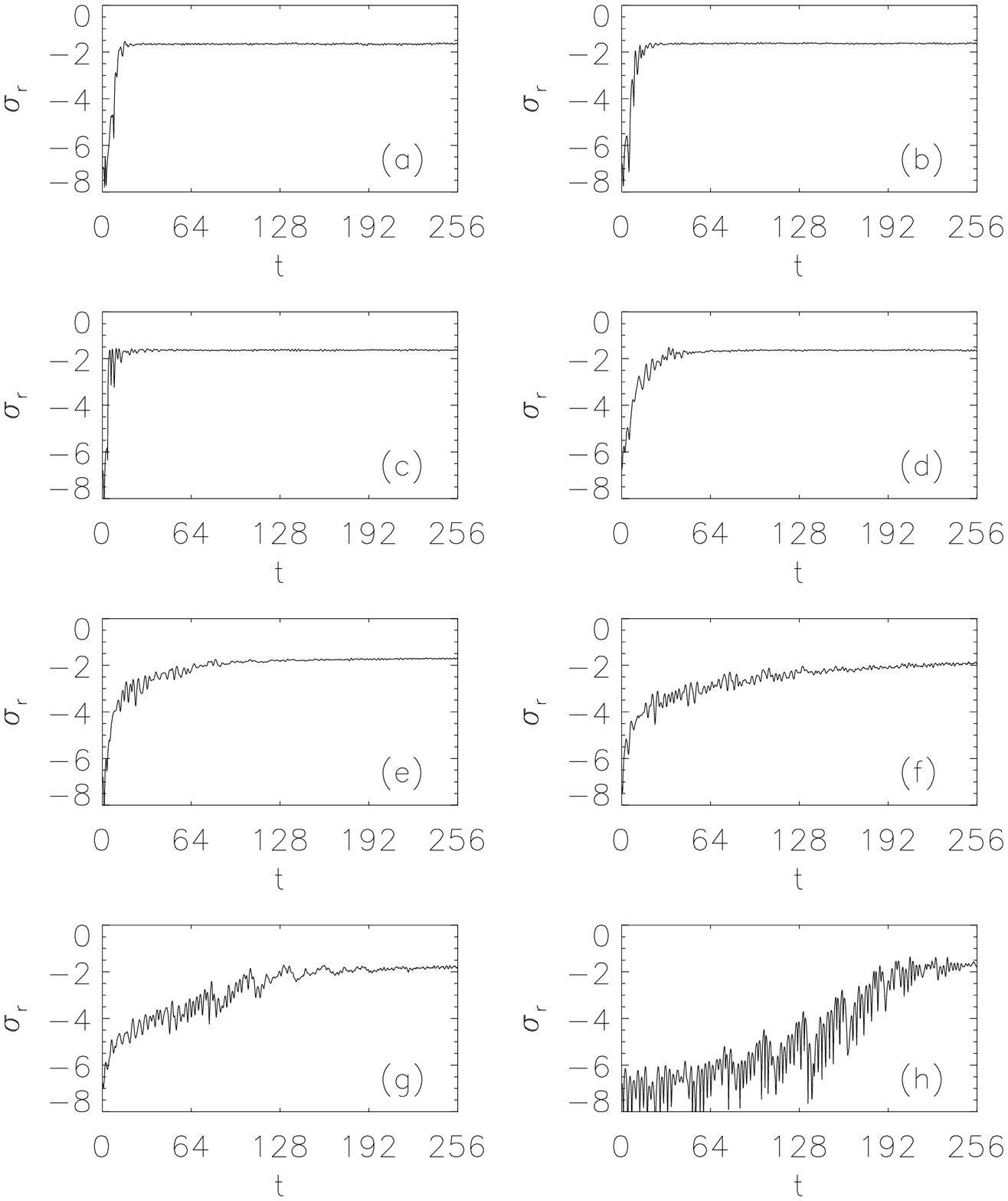}
           }
        \begin{minipage}{10cm}
        \end{minipage}
        \vskip -0.2in\hskip -0.0in
\caption{The configuration dispersion ${\sigma}_{r}$ associated with an
initially localised ensemble of `sticky' chaotic orbits evolved in frozen-$N$ 
realisations of the nonintegrable potential (2.6) for variable $N$. (a) 
$N=10^{2.5}$. (b) $N=10^{3}$. (c) $N=10^{3.5}$. (d) $N=10^{4}$. (e) 
$N=10^{4.5}$. (f) $N=10^{5}$. (g) $N=10^{5.5}$. (h) Evolution in the smooth 
potential.}
\vspace{-0.0cm}
\end{figure}
\vfill\eject
\begin{figure}[t]
\centering
\centerline{
        \epsfxsize=8cm
        \epsffile{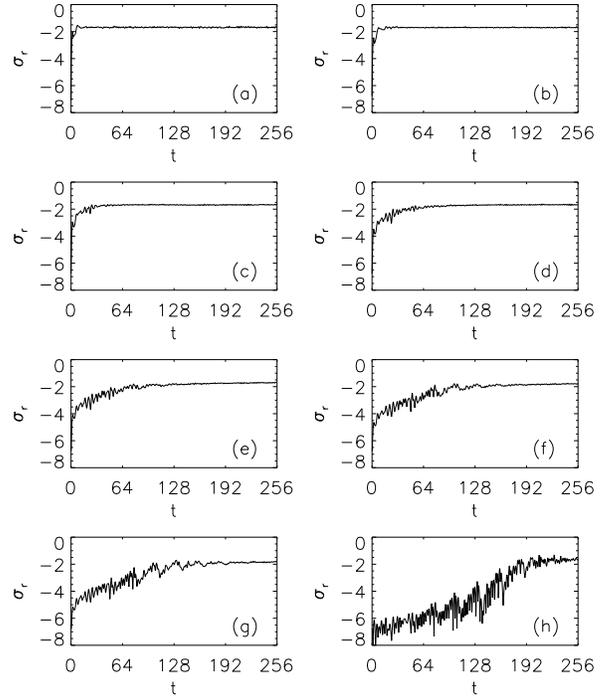}
           }
        \begin{minipage}{10cm}
        \end{minipage}
        \vskip -0.2in\hskip -0.0in
\caption{The configuration dispersion ${\sigma}_{r}$ associated with an
initially localised ensemble of `sticky' chaotic orbits evolved in the
nonintegrable potential (2.6) but perturbed by noise with ${\Theta}=1.0$ and 
variable ${\eta}$. 
(a) ${\eta}=10^{-2}$. (b) ${\eta}=10^{-2.5}$. (c) ${\eta}=10^{-3}$.
(d) ${\eta}=10^{-3.5}$. (e) ${\eta}=10^{-4}$. (f) ${\eta}=10^{-4.5}$.
(g) ${\eta}=10^{-5}$, (h) ${\eta}=10^{-7.5}$.}
\vspace{-0.0cm}
\end{figure}

\begin{figure}[t]
\centering
\centerline{
        \epsfxsize=8cm
        \epsffile{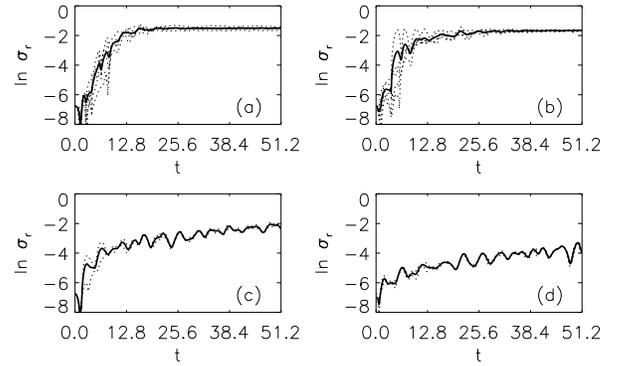}
           }
        \begin{minipage}{10cm}
        \end{minipage}
        \vskip -0.2in\hskip -0.0in
\caption{The configuration dispersion ${\sigma}_{r}$ associated with the 
ensemble exhibited in the preceding FIGURE, now restricted to a shorter 
time interval.
(a) $N=10^{2.5}$. (b) $N=10^{3.5}$. (c) $N=10^{4.5}$. (d) $N=10^{5.5}$.}
\vspace{-0.0cm}
\end{figure}

\pagestyle{empty}
\begin{figure}[t]
\centering
\centerline{
        \epsfxsize=8cm
        \epsffile{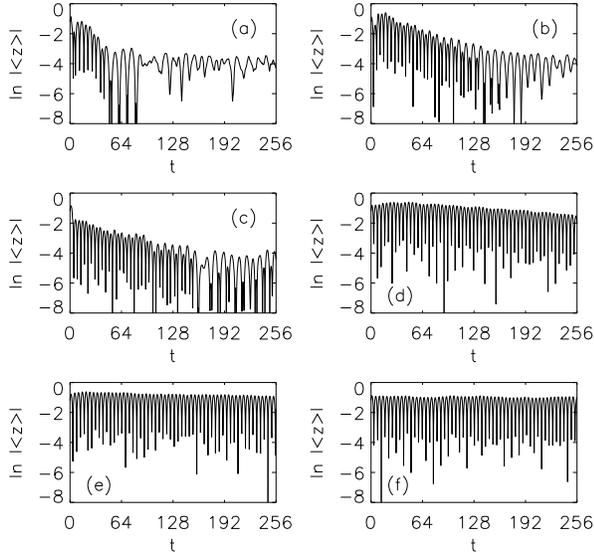}
           }
        \begin{minipage}{10cm}
        \end{minipage}
        \vskip -0.0in\hskip -0.0in
\caption{The quantity $\ln |{\langle}z{\rangle}|$  for an
initially localised ensemble evolved in frozen-$N$ realisations of the 
integrable ellipsoid potential (2.5) for variable $N$. (a) $N=10^{2.5}$. (b) 
$N=10^{3}$. (c) $N=10^{3.5}$. (d) $N=10^{4}$. (e) $N=10^{4.5}$. (f) 
$N=10^{5}$.}
\vspace{-0.0cm}
\end{figure}
\vfill\eject
\pagestyle{empty}
\begin{figure}[t]
\centering
\centerline{
        \epsfxsize=8cm
        \epsffile{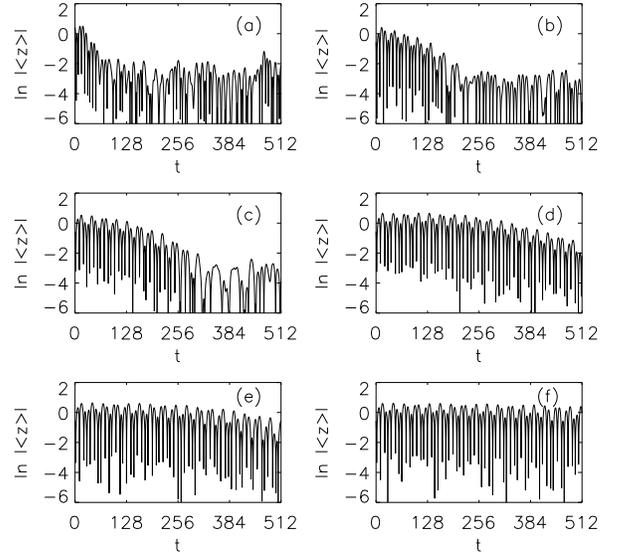}
           }
        \begin{minipage}{10cm}
        \end{minipage}
        \vskip -0.0in\hskip -0.0in
\caption{The quantity $\ln |{\langle}z{\rangle}|$  for an
initially localised ensemble evolved in frozen-$N$ realisations of the 
integrable Plummer potential (2.2) for variable $N$. (a) $N=1000$. 
(b) $N=10^{3.5}$. (c) $N=10^{4}$. (d) $N=10^{4.5}$. (e) $N=10^{5}$. (f) 
Evolution in the smooth potential.}
\vspace{-0.0cm}
\end{figure}
\vfill\eject
\pagestyle{empty}
\begin{figure}[t]
\centering
\centerline{
        \epsfxsize=8cm
        \epsffile{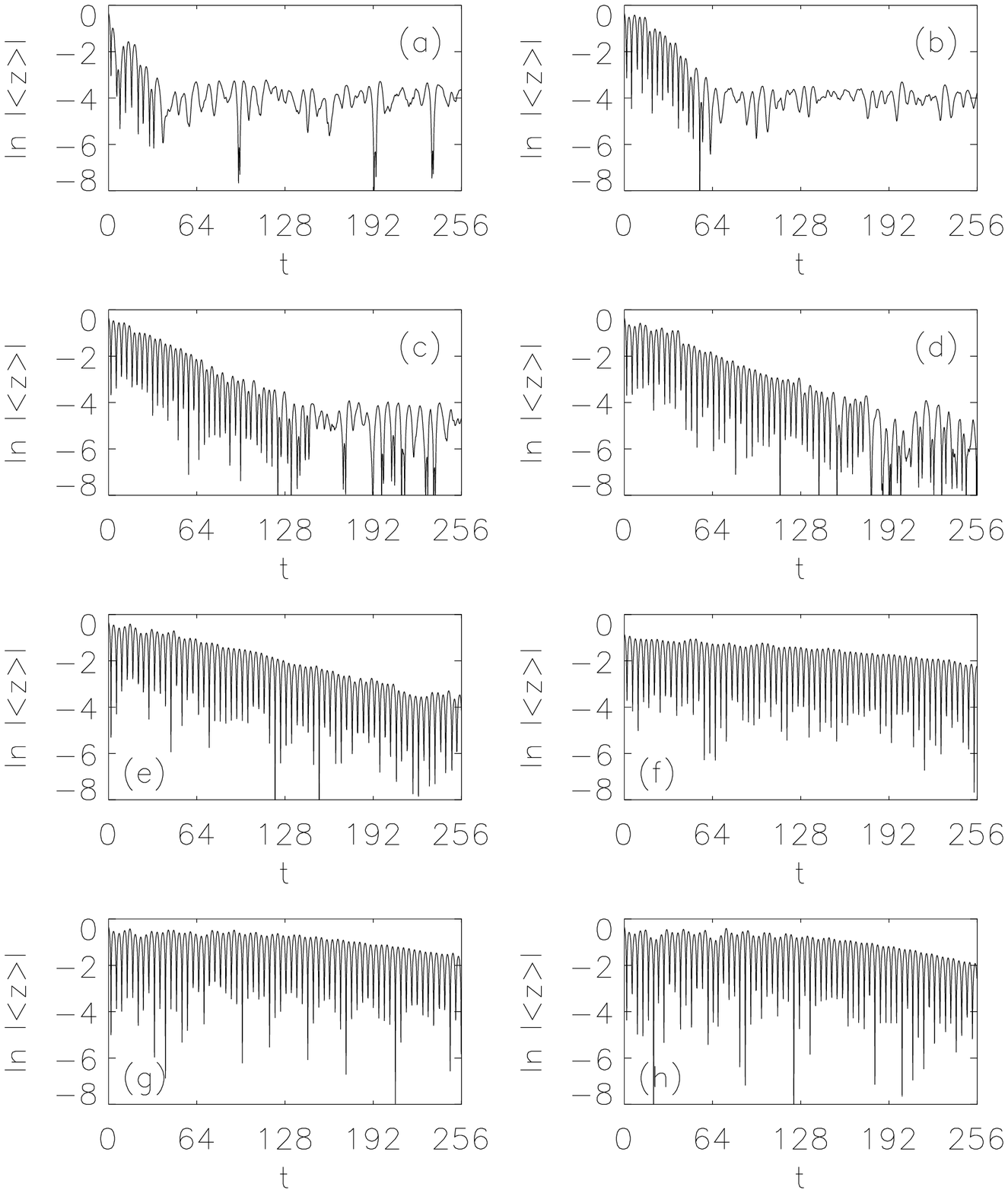}
           }
        \begin{minipage}{10cm}
        \end{minipage}
        \vskip -0.2in\hskip -0.0in
\caption{The quantity $\ln |{\langle}z{\rangle}|$  for an
initially localised ensemble of wildly chaotic orbits evolved in frozen-$N$ 
realisations of the nonintegrable potential (2.6) for variable $N$. (a) 
$N=10^{2.5}$. (b) $N=10^{3}$. (c) $N=10^{3.5}$. (d) $N=10^{4}$. 
(e) $N=10^{4.5}$. (f) $N=10^{5}$. (g) $N=10^{5.5}$. (g) Evolution in the 
smooth potential.}
\vspace{-0.0cm}
\end{figure}
\vfill\eject

\pagestyle{empty}
\begin{figure}[t]
\centering
\centerline{
        \epsfxsize=8cm
        \epsffile{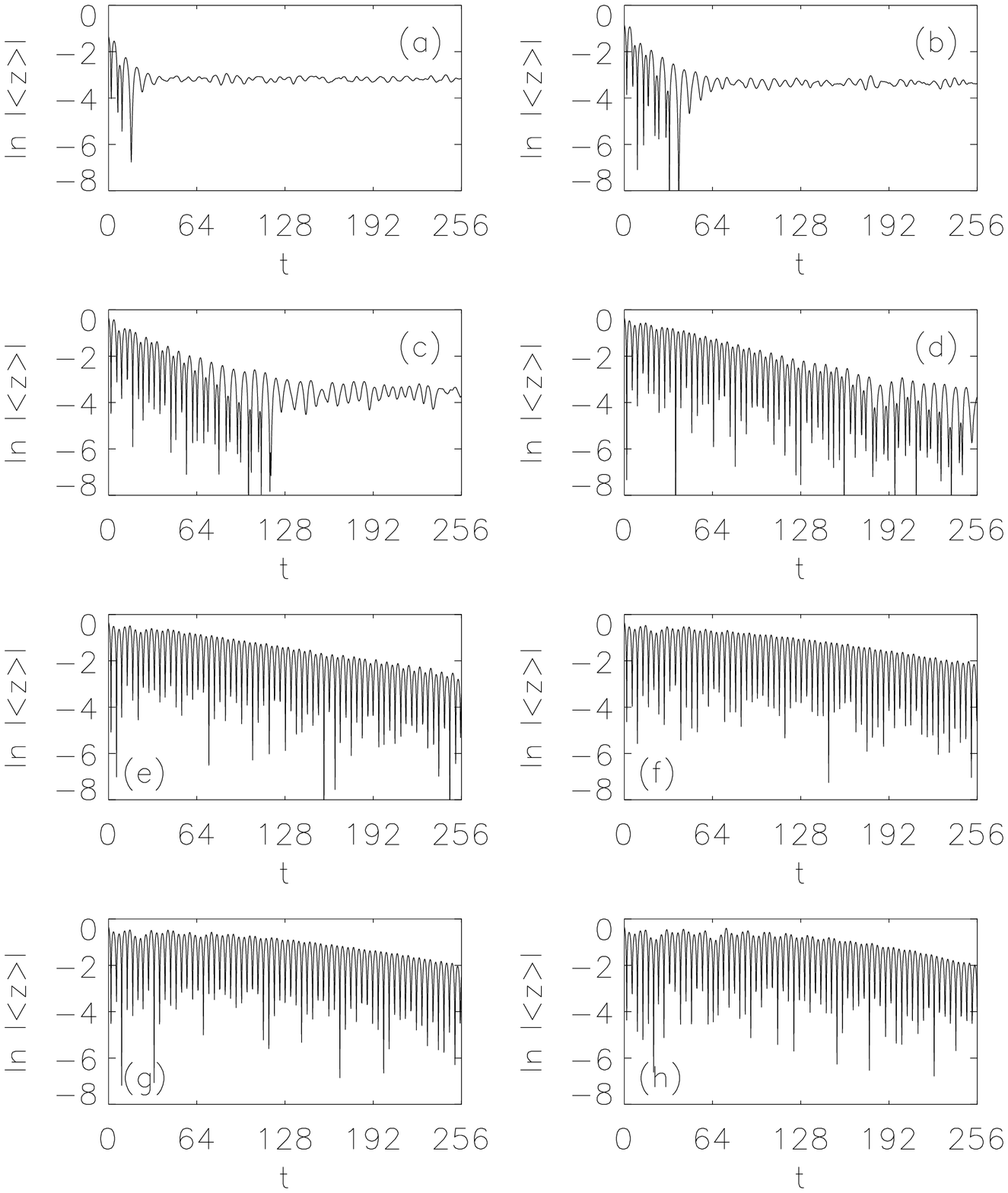}
           }
        \begin{minipage}{10cm}
        \end{minipage}
        \vskip -0.2in\hskip -0.0in
\caption{The quantity $\ln |{\langle}z{\rangle}|$  for an
initially localised ensemble of wildly chaotic orbits evolved in the
nonintegrable potential (2.6) but perturbed by noise with ${\Theta}=1.0$ and 
variable ${\eta}$. 
(a) ${\eta}=10^{-2}$. (b) ${\eta}=10^{-2.5}$. (c) ${\eta}=10^{-3}$.
(d) ${\eta}=10^{-3.5}$. (e) ${\eta}=10^{-4}$. (f) ${\eta}=10^{-4.5}$.
(g) ${\eta}=10^{-5}$, (h) ${\eta}=10^{-7.5}$.}
\vspace{-0.0cm}
\end{figure}
\vfill\eject
\pagestyle{empty}
\begin{figure}[t]
\centering
\centerline{
        \epsfxsize=8cm
        \epsffile{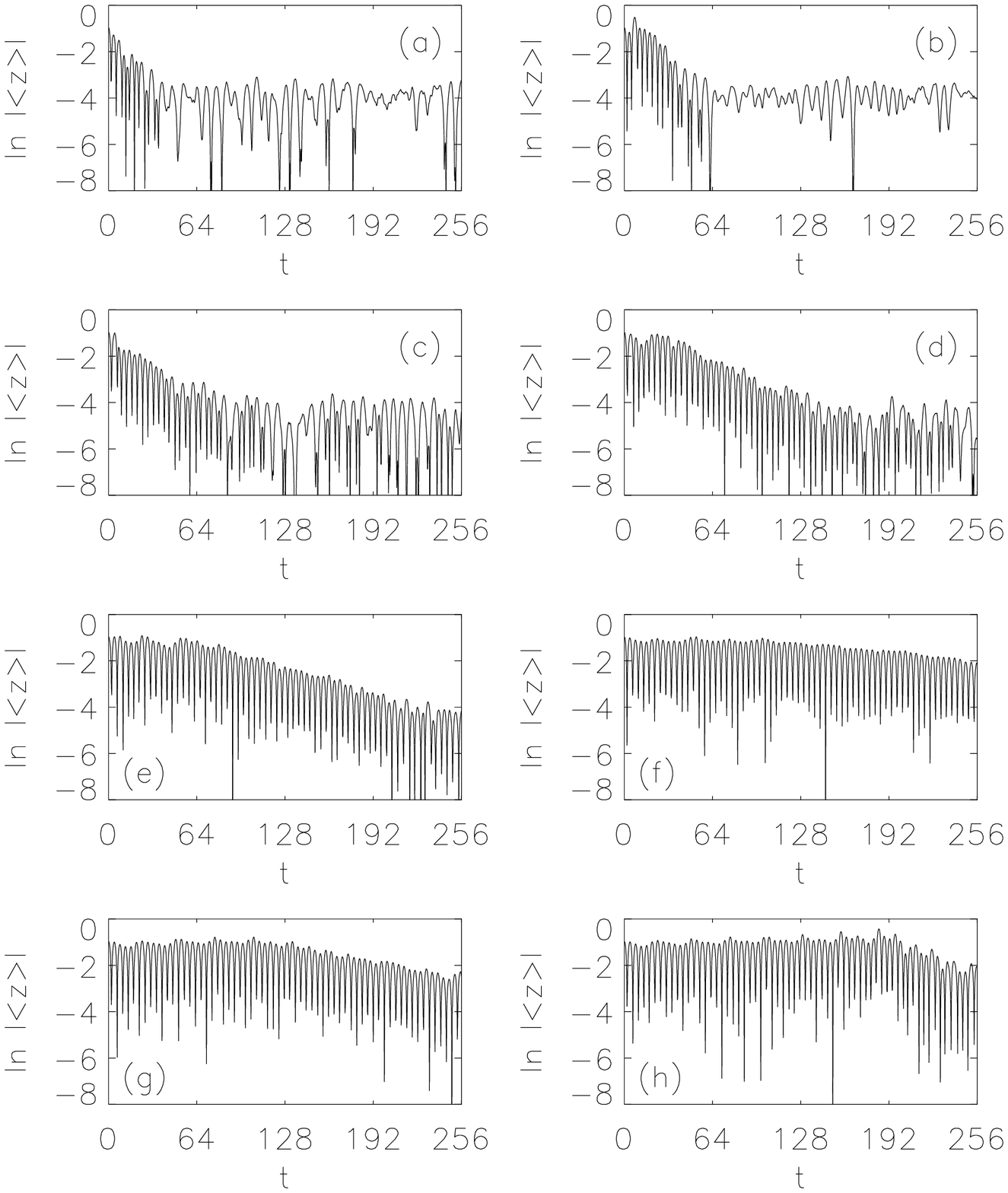}
           }
        \begin{minipage}{10cm}
        \end{minipage}
        \vskip -0.2in\hskip -0.0in
\caption{The quantity $\ln |{\langle}z{\rangle}|$  for an
initially localised ensemble of `sticky' chaotic orbits evolved in frozen-$N$ 
realisations of the nonintegrable potential (2.6) for variable $N$. (a) 
$N=10^{2.5}$. (b) $N=10^{3}$. (c) $N=10^{3.5}$. (d) $N=10^{4}$. (e) 
$N=10^{4.5}$. (f) $N=10^{5}$. (g) $N=10^{5.5}$. (g) Evolution in the smooth 
potential.}
\vspace{-0.0cm}
\end{figure}
\vfill\eject

\pagestyle{empty}
\begin{figure}[t]
\centering
\centerline{
        \epsfxsize=8cm
        \epsffile{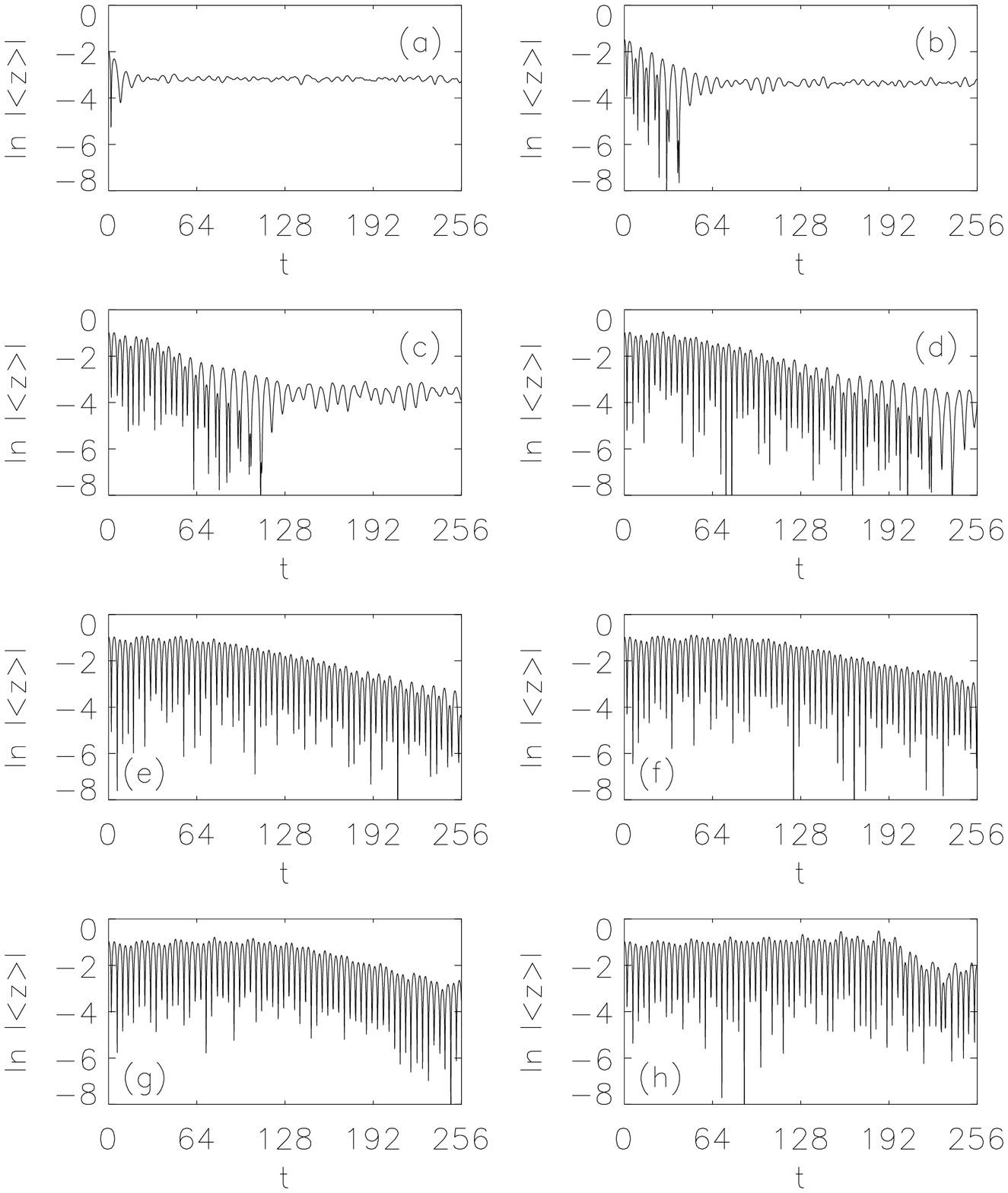}
           }
        \begin{minipage}{10cm}
        \end{minipage}
        \vskip -0.2in\hskip -0.0in
\caption{The quantity $\ln |{\langle}z{\rangle}|$  for an
initially localised ensemble of `sticky' chaotic orbits evolved in the
nonintegrable potential (2.6) but perturbed by noise with ${\Theta}=1.0$ and 
variable ${\eta}$. 
(a) ${\eta}=10^{-2}$. (b) ${\eta}=10^{-2.5}$. (c) ${\eta}=10^{-3}$.
(d) ${\eta}=10^{-3.5}$. (e) ${\eta}=10^{-4}$. (f) ${\eta}=10^{-4.5}$.
(g) ${\eta}=10^{-5}$, (h) ${\eta}=10^{-7.5}$.}
\vspace{-0.0cm}
\end{figure}
\vfill\eject
\pagestyle{empty}
\begin{figure}[t]
\centering
\centerline{
        \epsfxsize=8cm
        \epsffile{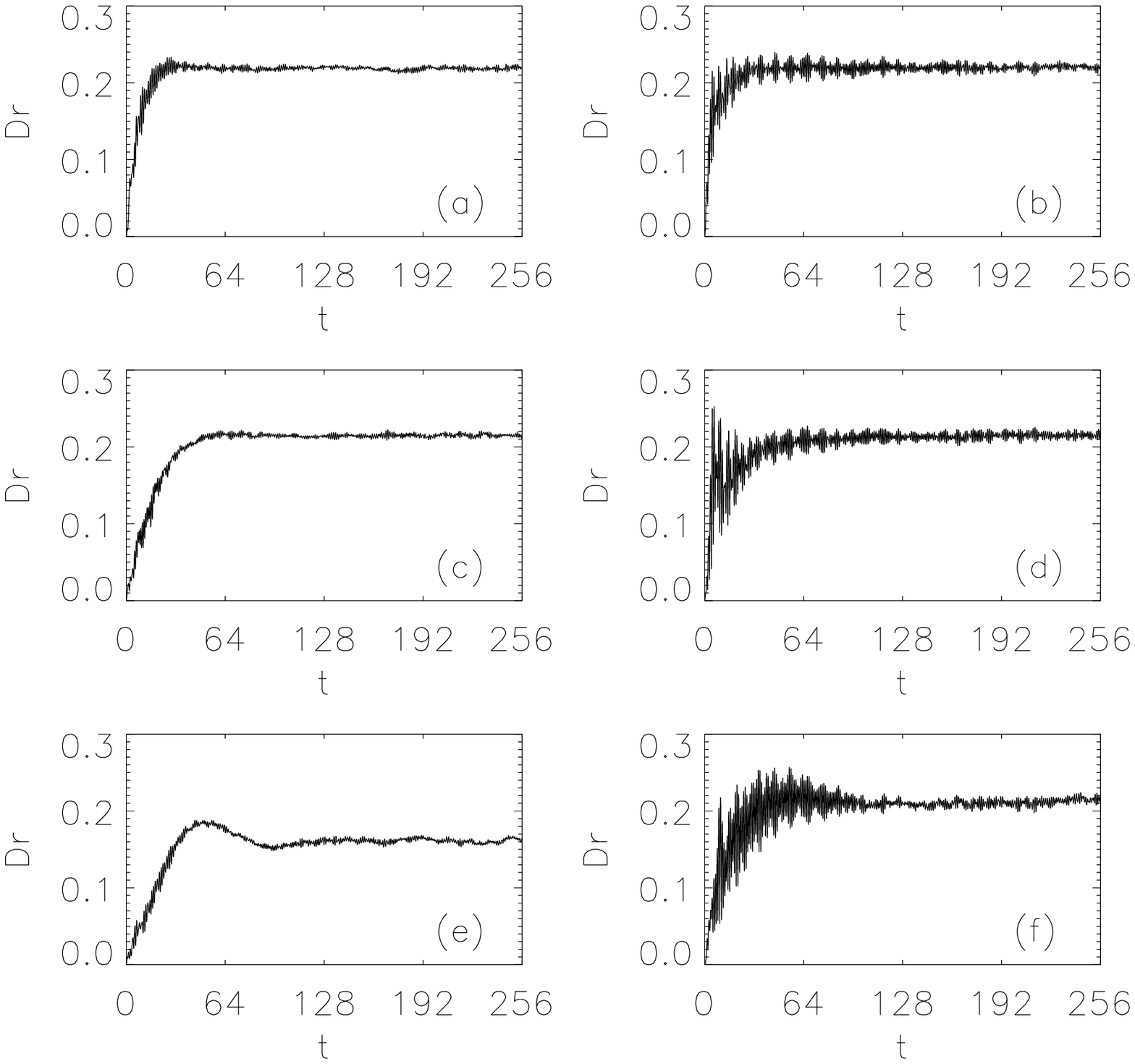}
           }
        \begin{minipage}{10cm}
        \end{minipage}
        \vskip -0.2in\hskip -0.0in
\caption{The quantity ${\cal D}r$  for two different initially localised 
ensembles of `sticky' chaotic orbits with the same (low) energies, each 
evolved in frozen-$N$ realisations of the Dehnen potential. 
(a) A `regular' ensemble with $N=10^{4}$. 
(b) A `chaotic' ensemble with $N=10^{4}$.
(c) The `regular' ensemble with $N=10^{4.5}$. 
(d) The `chaotic' ensemble with $N=10^{4.5}$.
(a) The `regular' ensemble with $N=10^{5}$. 
(b) The `chaotic' ensemble with $N=10^{5}$.}
\vspace{-0.0cm}
\end{figure}

\pagestyle{empty}
\begin{figure}[t]
\centering
\centerline{
        \epsfxsize=8cm
        \epsffile{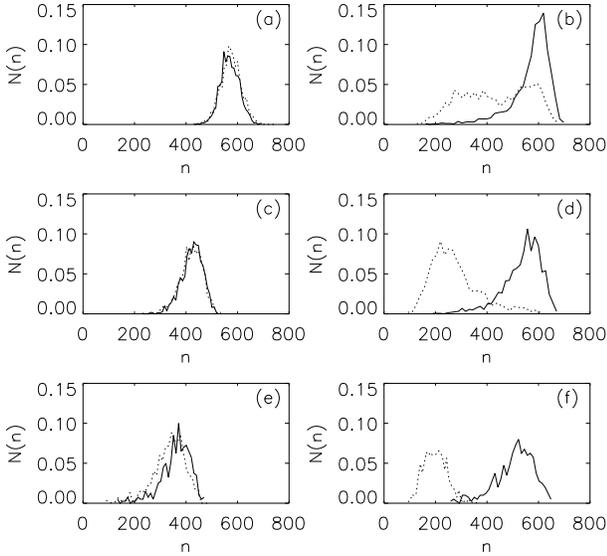}
           }
        \begin{minipage}{10cm}
        \end{minipage}
        \vskip -0.2in\hskip -0.0in
\caption{The distribution of complexities, $N(n)$,  for two different 
initially localised 
ensembles of `sticky' chaotic orbits with the same (low) energies, each 
evolved in frozen-$N$ realisations of the Dehnen potential. 
(a) A `regular' ensemble with $N=10^{4}$. 
(b) A `chaotic' ensemble with $N=10^{4}$.
(c) The `regular' ensemble with $N=10^{4.5}$. 
(d) The `chaotic' ensemble with $N=10^{4.5}$.
(e) The `regular' ensemble with $N=10^{5}$. 
(f) The `chaotic' ensemble with $N=10^{5}$.}
\vspace{-0.0cm}
\end{figure}

\pagestyle{empty}
\begin{figure}[t]
\centering
\centerline{
        \epsfxsize=8cm
        \epsffile{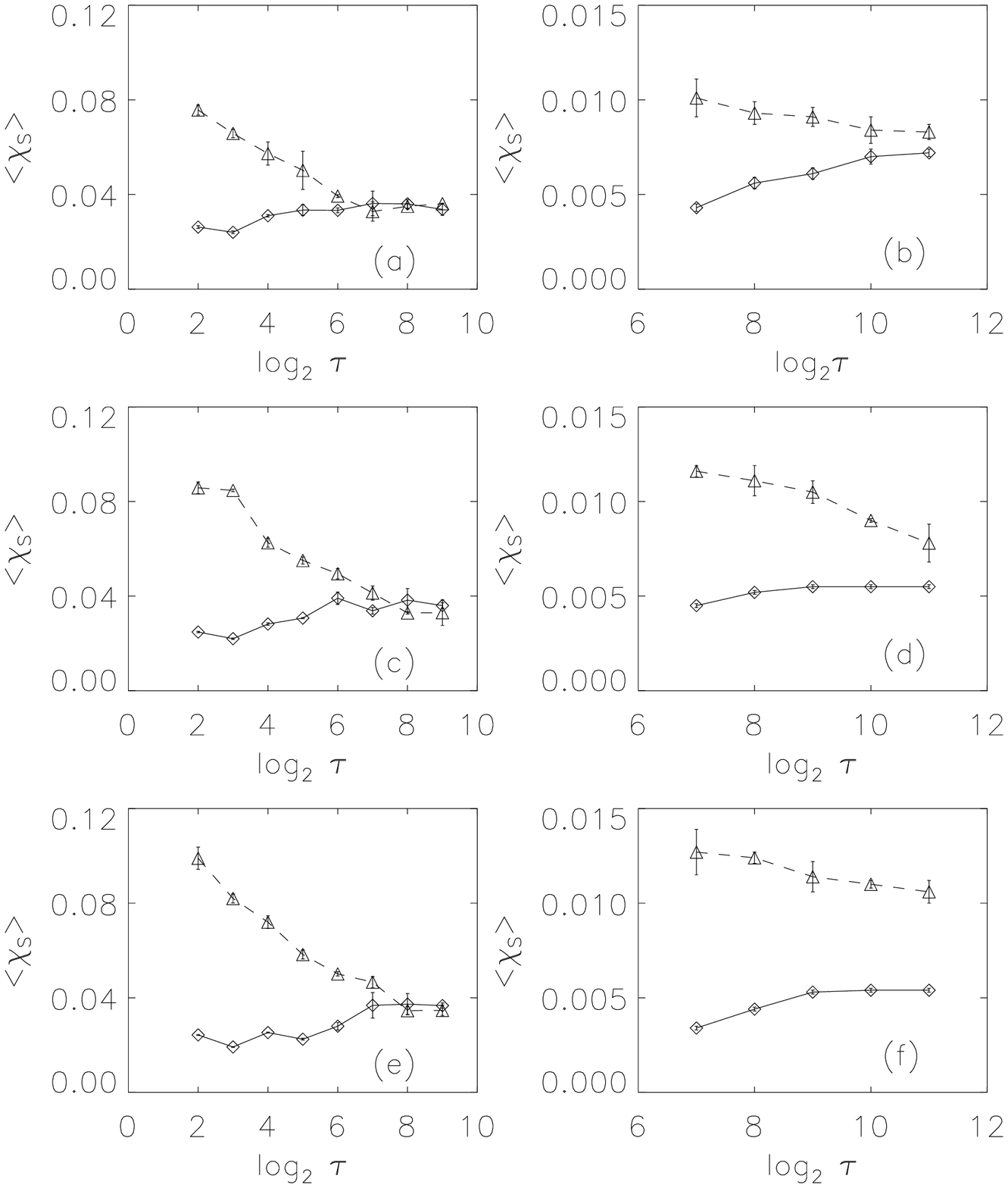}
           }
        \begin{minipage}{10cm}
        \end{minipage}
        \vskip -0.2in\hskip -0.0in
\caption{The mean Lyapunov exponent ${\langle}{\chi}_{S}{\rangle}$, for 
ensembles evolved in the smooth Dehnen potential, selecting as initial 
conditions the final phase space coordinates of orbits that had been evolved 
in frozen-$N$ potentials for some time ${\tau}$. 
(a) Initially regular
(solid line) and chaotic (dashed line) low energy orbits evolved with 
$N=10^{4}$. (b) Initially regular (solid line) and chaotic (dashed line) 
higher energy orbits evolved with $N=10^{4}$. (c) The same as (a) but for
$N=10^{4.5}$. (d) The same as (b) but for $N=10^{4.5}$. (e) The same as (a) 
but for $N=10^{5}$. (f) The same as (b) but for $N=10^{5}$.}
\vspace{-0.0cm}
\end{figure}

\begin{references}
\bibitem{Sh2}These ensembles sampled Merritt and Fridman's shell 8, for which
$E{\;}{\approx}{\;}-0.4478$.
\bibitem{mod}Modulo the case of a constant density sphere, where evey
unperturbed orbit has the same unperturbed frequencies, so that linear phase 
mixing is impossible.
\bibitem{Sh1} The ensembles sampled the lowest energy `shell' considered by
Merritt and Fridman\cite{MF}, for which, in the continuum limit, 
$E{\;}{\approx}{\;}-0.9964$.
\bibitem{st}One might, naively, suppose that the relative number of regular
and chaotic orbits in each ensemble would provide a more satisfactory probe. 
However, as discussed in \cite{Siop}, `stickiness' is so pronounced in the
Dehnen potential that integrations for times as short as $t=10^{3}-10^{4}$
are not always sufficient to make sharp distinctions between regularity and
chaos. For the lower energy orbits, ${\chi}_{S}$ was computed by integrating
for a time $T=256$; for the higher energy, $T=1024$.
\end{references}
\end{document}